\documentclass[prd,aps,10pt,nofootinbib]{revtex4-1}
\usepackage[dvips]{graphicx,color}
\usepackage{graphics}
\usepackage{pstricks}
\usepackage{epstopdf}
\usepackage[T1]{fontenc}
\usepackage{amsmath}
\usepackage{amsfonts}
\usepackage{amssymb}
\usepackage{latexsym}
\newcommand{\ba}{\begin{eqnarray}}
\newcommand{\ea}{\end{eqnarray}}
\newcommand{\be}{\begin{equation}}
\newcommand{\ee}{\end{equation}}
\newcommand{\bc}{\begin{center}}
\newcommand{\ec}{\end{center}}
\newcommand{\n}{\nonumber\\}
\begin{document}
\title{Holographic thermalization of charged operators}
\author{Alejandro Giordano}
\affiliation{Instituto de F\'\i sica de La Plata - CONICET \&
Departamento de F\'isica - UNLP,\\ C.C. 67, 1900 La Plata,
Argentina}
\author{Nicol\'as Grandi}
\affiliation{Instituto de F\'\i sica de La Plata - CONICET \&
Departamento de F\'isica - UNLP,\\ C.C. 67, 1900 La Plata,
Argentina\\Abdus Salam International Centre for Theoretical Physics,
Associate Scheme \\ Strada Costiera 11, 34151, Trieste, Italy}
\author{Guillermo Silva}
\affiliation{Instituto de F\'\i sica de La Plata - CONICET \&
Departamento de F\'isica - UNLP,\\ C.C. 67, 1900 La Plata,
Argentina\\Abdus Salam International Centre for Theoretical Physics,
Associate Scheme \\ Strada Costiera 11, 34151, Trieste, Italy}
\begin{abstract}
We study a light-like charged collapsing shell in AdS-Reissner-Nordstrom spacetime,
investigating whether the corresponding Vaidya metric is supported
by matter that satisfies the null energy condition.  We find that,
if the absolute value of the charge decreases during the collapse,
energy conditions are fulfilled everywhere in spacetime. On the
other hand, if the absolute value of the charge increases, the
metric does not satisfy energy conditions in the IR region.
Therefore, from the gauge/gravity perspective, this last case is
only useful to study the thermalization of the UV degrees of
freedom. For all these geometries, we probe the thermalization
process with two point correlators of charged operators, finding
that the thermalization time grows with the charge of the operator, as well as with the dimension of space.
\end{abstract}
\maketitle
%
%
\section{Introduction}
%
%
The study of non-equilibrium processes in quantum field theory is an
important and active area of research (see \cite{noneq} for
references). The subject has most diverse applications that range
from astrophysics and cosmology, {\em i.e.} the study of particle
production at the end of inflation and the generation of density
fluctuations, to relativistic heavy ion collisions, {\em i.e.} the
dynamics of the quark-gluon plasma at RHIC. For the case of
near-equilibrium states there exist well developed formalisms:
linear response, kinetic theory or fluid dynamics. This techniques
rely on the existence of a dilute equilibrium ensemble  or weak
coupling. However in many physical instances the system to be
studied can be strongly coupled or in a dense state or both, and
therefore in these situations the standard techniques do not apply.

The question of whether a given subset of the degrees of freedom of
a certain system reaches equilibrium starting from an arbitrary
initial state, and whether such equilibrium can be described by a
thermal density matrix, is an open one whose answer seems to depend
on the details of the underlying dynamics \cite{nonequil2}. A large
number of numerical and analytical research on such {\it quenching}
process has been published (see \cite{quenches} and references
therein), regarding both weakly coupled systems in the perturbative
regime \cite{nonequil-peturbative}, and  integrable systems
\cite{nonequil-integrable}. Moreover, strong coupling QFT's have
been studied numerically in the lattice \cite{nonequil-lattice}.

In this paper we will consider the study of far from equilibrium
strongly coupled systems using the gauge/gravity correspondence. As
it is well known, the holographic correspondence is the natural
arena for analyzing strongly coupled QFT problems translating them
into classical gravity computations. For near-equilibrium states,
the correspondence has been extensively studied with particular
emphasis in the linear response and the hydrodynamic regime, in
terms of perturbations of the black hole geometry (see the recent  reviews
\cite{hube,HotQCD}). More recently, attention has been paid to
far-from-equilibrium states. Reference \cite{bala} made a
holographic proposal to model the sudden injection of energy into
the QFT vacuum state, and its subsequent thermalization, in terms of
an AdS Vaidya geometry. The {\it dynamical} Vaidya
spacetime physically corresponds to the collapse of a homogeneous
massless shell in AdS, leading to the formation of a black hole \cite{Vaidya1}. It
interpolates between pure AdS spacetime (vacuum) in the distant past
to an AdS black hole (thermal state) in the distant future. Two
point functions of operators $\cal O$ of large conformal dimensions,
which in the dual picture can be computed in the semiclassical
approximation in terms of geodesics, and Wilson loops and
entanglement entropy, which in the gravity perspective relate to
minimal (hyper)surfaces, were used as probes of thermalization. The
resulting picture is that the UV degrees of freedom thermalize
first, followed later by the IR ones (top-down thermalization
\cite{bala}). From the gravity perspective this is
an expected result since IR probes 
explore deeply in the radial direction being therefore sensitive to
the shell position for longer times.

The aforementioned results were generalized in \cite{martin,subad}
to situations with a non-trivial chemical potential or, in the dual
perspective, to the case in which the system experiences a sudden
injection of both energy and particles. The dual geometry was chosen
to be that of a charged collapsing null shell interpolating between
AdS in the distant past and asymptotically AdS Reissner-Nordstrom
black hole (AdSRN) in the distant future. The thermalization was
probed by two point functions of chargeless operators, Wilson loops,
and entanglement entropy. The emergent picture of \cite{martin} was
that as the final chemical potential is increased, it takes longer for the
system  to thermalize (see \cite{Garfinkle}-\cite{elena2} for
related works).

It is the aim of the present paper to extend the studies mentioned
above. In particular, since the Vaidya geometry is known to be
supported by an energy-momentum tensor satisfying null energy
conditions, in the present work we analyze the energy conditions of
the previously explored charged  Vaidya geometries
\cite{martin,subad} and its further generalizations. We afterwards
study the thermalization process by probing the system with charged
operators.

The organization of the paper is the following: In section
\ref{background} we analyze whether the Vaidya metric used in
\cite{martin,subad}, that represents a quench on the chemical
potential and temperature in the dual field theory, is supported by
matter satisfying null energy conditions, in section \ref{probes} we
probe the thermalization process with two point functions of charged
operators. The results are presented and discussed in section
\ref{results}. We conclude with a summary in section
\ref{conclusion}. In the appendices we discuss
Eddingtong-Finkelstein coordinates, the worldline formalism and the
WKB approximation, used in the text to obtain the two-point
functions of charged operators.
%
%
%
%
\section{Background geometry}
\label{background}
%
%
The gauge/gravity duality relates a strongly coupled quantum field
theory (QFT) in $d$ flat spacetime dimensions with a weakly coupled
gravity theory in a $d+1$ dimensional spacetime which asymptotes to
anti-de Sitter spacetime. According to the standard gauge/gravity
dictionary, a finite temperature state of the field theory is
represented by a geometry with a horizon in the bulk side. Moreover,
a global symmetry in the field theory induces a gauge symmetry in
the gravity side. As a consequence, the presence of a chemical
potential for a global $U(1)$ charge at finite temperature on the
QFT side is described in the dual gravitational picture by the
presence of an electrically charged black hole
[\onlinecite{hartnoll}]. In the simplest example, an equilibrium
state with finite temperature and chemical potential is represented
by an AdSRN black hole with given mass and charge. On the other
hand, a process in which the temperature and chemical potential vary
can be represented by a metric which interpolates between two such
geometries with different values of mass and charge. In this
section, we describe those geometries and analyze the energy
momentum tensor that is needed to support them. We model the
dynamical geometry by the collapse of a thin shell of null dust. It
turns out to be convenient to substitute the standard time
coordinate $t$, which is not constant across the shell, by an
infalling radial null coordinate $v$ which is.

In what follows, we take $d$ to be the spacetime dimension of the
dual field theory, hence our bulk geometry will be $d+1$
dimensional. The indices $\mu,\nu=0\dots d$ denote the bulk
coordinates $x^\mu=(v,{\bf x},z)$.
%
%
\subsection{Equilibrium state}
\label{thermal}
\subsubsection*{Bulk geometry}
\label{thermal-bulk}
%
%
In Eddington-Finkelstein ingoing null coordinates, the metric and
gauge fields corresponding to a planar AdS$_{d+1}$
Reissner-Nordstrom black hole take the form (see appendix \ref{EF})
\ba ds^2&=&\frac{L^2}{z^2}\left(-f dv^2-2dv dz+d{\bf
x}^2_{d-1}\right)\,, \label{RNADSBH}
\\
F&=&  L\, F_{zv}\, dz \wedge dv\,,
\label{RNADSBHA} \ea
where $f$ and $F_{zv}$ are functions of $z$ that read
\ba f&=&1 -M z^d+Q^2 z^{2(d-1)}\,, \label{soluf}
\\
F_{zv}&=&-\gamma (d\!-\!2){Q}
z^{d-3},~~~~~~~~~~~~\mbox{with}~~~~~\gamma^2=\frac{d-1}{2(d-2)}\,.
\label{soluphi} \ea
The static background \eqref{RNADSBH}-\eqref{RNADSBHA} is a vacuum
solution of the $(d+1)$-dimensional Einstein-Maxwell system with
negative cosmological constant $\Lambda=-{d(d-1)}/{2(\kappa L)^2}$
\be
S_{EM}=\frac1{2\kappa^2}\int
d^{d+1}x\sqrt{-g}\left[R+\frac{d(d-1)}{L^2}\right]-\frac1{4 }\int
d^{d+1}x\sqrt{-g} F_{\mu\nu}F^{\mu\nu}\,.
\ee
Setting $\kappa^2 =1$
the equations of motion following from $S_{EM}$ are
\ba
R_{\mu\nu}-\frac12g_{\mu\nu}R+\Lambda g_{\mu\nu}&=&
T_{\mu\nu}^{\sf Maxwell},
\label{eim}\\
\nabla_\mu F^{\mu\nu}&=&0\,, \label{einstmax}
\ea
where
\be
T_{\mu\nu}^{\sf
Maxwell}=F_{\mu\alpha}F_{\nu\beta}g^{\alpha\beta}-\frac14g_{\mu\nu}F^2\,.
\ee

The geometry \eqref{RNADSBH} asymptotes AdS space with radius $L$ as
we approach the boundary located at $z= 0$, and the constants $M$
and $Q$ correspond to the ADM mass and electric charge respectively.
The metric \eqref{RNADSBH} has a curvature singularity at $z\to
\infty$ and has horizons whenever the function $f$ vanishes. To
characterize the horizons, notice that $f$ has two stationary
points, one at  $z=0$ at which $f=1$, and a second one (a local
minimum) at $z_{min}=(d\, M /2(d-1)Q^2)^{1/(d-2)}$. The curvature
singularity will be hidden from the outside as long as
$f(z_{min})\le0$, which implies a constraint on the possible values
of $M,Q$
\be
\left(\frac{M^{d-1}}{Q^d}\right)^2\ge\left(\frac {2(d-1)}d
\right)^d\left(\frac{2(d-1)}{d-2}\right)^{d-2} \,.
\label{hori}
\ee
Whenever this inequality is satisfied, we generically have two
horizons at $z=z_\pm$ (inner/outer). Moreover, when the bound is
saturated the two horizons coincide and the configuration is called
an {\it extremal black hole solution} (more on this below). We will
demand the constraint \eqref{hori} on all our solutions in order to
have a physically sensible gravitational background.

Although charged black holes generically depend on two arbitrary
parameters $Q$ and $M$, this is not so in the planar horizon case.
The absence of a scale on the horizon geometry allows us to get rid
of one of the parameters. Explicitly, the rescaling $(v,{\bf x},
z)=z_- (\tilde v,\tilde{\bf x}, \tilde z)$ maps the (outer) horizon
position to $\tilde z=1$. Defining  $\tilde M=Mz_-^d$ and $\tilde
Q=Qz_-^{d-1}$ one finds that $\tilde M=1+\tilde Q^2$ resulting into
\cite{hartnoll}
\ba
f&=&1-(1+\tilde Q^2) \tilde z^d+\tilde Q^2 \tilde z^{2(d-1)}\,,
\label{geoBH}\\
F_{zv}&=&-\gamma\,(d\!-\!2){\tilde Q} \tilde z^{d-3}\, .
\label{geoA}
\ea
This is the parametrization often used in the literature for the
planar Reissner-Nordstrom AdS black hole. It automatically satisfies
the constraint \eqref{hori}, and notice that $\tilde Q$ can take any
arbitrary value. The geometry  \eqref{geoBH} has generically two
horizons. Depending on the value of $\tilde Q$ one has: (i) an outer
one at $\tilde z_-=1$ and an inner one at $\tilde z_+>1$ for $
|\tilde Q|\le\sqrt{d/{(d-2)}}$, (ii) coincident horizons at $\tilde
z_\pm=1$ for $\tilde Q^2=d/(d-2)$ (extremal BH), and (iii)  an inner
horizon at $\tilde z_+=1$ and an outer one at $\tilde z_-<1$ for
$|\tilde Q|>\sqrt{d/{(d-2)}}$. An appropriate  rescaling of the
holographic coordinate maps the type (iii) solutions to the type (i)
ones.

Summarizing, the background \eqref{geoBH}-\eqref{geoA} with $|\tilde
Q|\le\sqrt{d/{(d-2)}}$ parametrizes the most general static planar
AdSRN solution with an outer horizon located at $\tilde z_-=1$ and
an inner horizon located at $\tilde z_+\ge 1$.
%
%
%
%
%
%
%
\subsubsection*{Boundary theory }
\label{static-boundary}
%
%
%
As mentioned above, the AdS boundary is located at $z = 0$ and, as
it is well known, the static geometry
\eqref{RNADSBH}-\eqref{RNADSBHA} represents a dual QFT  equilibrium
state characterized by a chemical potential $\mu$ and a temperature
$\sf T$ \cite{hartnoll}. The bulk variable $v$ is identified with
the dual gauge theory time $t$ since both coincide at $z=0$ (see
\eqref{time} below).

The standard procedure to relate the boundary and bulk parameters is
to impose regularity of the Wick rotated solution.  For the static
case \eqref{soluf}-\eqref{soluphi}, the redefinition
\be -i dt_E=dv+\frac{dz}{f}\,, \label{time} \ee
turns the geometry \eqref{RNADSBH} into the Euclidean form
\ba ds^2&=&\frac{1}{z^2}\left(f dt_E^2+\frac{dz^2}{f}+d{\bf
x}^2_{d-1}\right)\,. \label{RNADSBH-st} \ea
We note in passing that if the function $f$ were $v$-dependent, as
it will become later when we define Vaidya metrics, the redefinition
\eqref{time} would not be allowed, the right hand side being not an
exact differential. The metric \eqref{RNADSBH-st} is regular at the
outer horizon  $z=z_-$ if we periodically identify $t_E\equiv t_E
+\beta$, with
\be \beta = \frac{4\pi}{|f_z(z_-)|}\,, \ee
the subindex $z$ on $f_z$ denotes derivative with respect to $z$.
The boundary theory temperature is then defined as
\be {\sf T}\equiv\frac1\beta=\frac{d\, M z_-^{d}
-2(d-1)Q^2z_-^{2d-2}}{4\pi z_-}\,. \label{temp} \ee
Regarding the chemical potential of the boundary theory, it relates
to the black hole charge as follows: we can choose the bulk gauge
potential to be
\be A_v = -\gamma   QL\, z^{d-2}+\mu L  \,, \ee
with $\mu$ arbitrary. From \eqref{time} one has $i A_{t_E} = -\gamma
QL z^{d-2}+\mu L$. Since the $t_E$ circle smoothly collapses at the
outer horizon, the gauge field must satisfy $A_{t_E}(z_-)=0$ to
avoid singularities.   This condition fixes the asymptotic value of
the gauge field to
\be \mu =\gamma Q z_-^{d-2}\,. \ee
For a fixed mass black hole, increasing the chemical potential ({\em
i.e.} the black hole charge) decreases the black hole temperature.
The extremal bound in \eqref{hori} corresponds to the ${\sf T}=0$
and $\mu\ne0$ case. In the gauge/gravity duality context one
identifies the gauge field boundary value $\mu$ as the source for
the QFT conserved charge operator.
%
%
%
%
%
%
\subsection{Time dependent states}
\label{timedep}
\subsubsection*{Bulk geometry}
\label{timedep-bulk}
%
%
%
%
To construct a time dependent geometry, we  promote the
$z$-dependent functions $f$ and $F_{zv}$ on \eqref{RNADSBHA} to
$(z,v)$-dependent functions. In other words, we keep the ansatz for
the metric and the field strength \eqref{RNADSBH}-\eqref{RNADSBHA},
but the functions $f,F_{zv}$ now depend on both variables $z,v$.
Under this conditions, extra matter contributions $T_{\mu\nu}^{\sf
Matter }$ and $j^\mu_{\sf Matter}$ need to be added to the right
hand side of \eqref{eim}-\eqref{einstmax}, in order to satisfy
Einstein-Maxwell equations of motion. They physically represent an
infalling charged matter shell giving birth to the black hole
[\onlinecite{Vaidya1}]. The additional contribution to the energy
momentum tensor is defined as
\be T_{\mu\nu}^{\sf Matter}=  \left[G_{\mu\nu}-\frac {(d - 1) d}{2
L^2}\, g_{\mu\nu}\right] - T_{\mu\nu}^{\sf Maxwell} \,. \ee
Working out the components of $T^{\sf Matter}_{\mu\nu}$ in terms of
the functions $f,F_{zv}$ and their derivatives one finds
\ba
 T_{vv}^{\sf Matter}&=& \frac{ (d-1)}{2z^2} \bigg( f\big(  \,z
f_z - d \,f+d \big)- z\,f_v\bigg) -\frac{z^2}{2}\, fF_{zv}^2 \,,
\label{eom}
\\
T_{vz}^{\sf Matter } &=& \frac{ (d-1)}{2z^2} \bigg( z\,f_z - d \,f +
d\bigg) -\frac{z^2}{2}\, F_{zv}^2\,, \label{eom2}
\\
T_{x_ix_j}^{\sf Matter }&=&\delta_{ij} \left[\frac{ 1}{2}
f_{zz}-\frac{ (d-1) }{2z^2} \bigg(
  2  z\,f_z - d\,f + d \bigg)  -  \frac{z^2}{2}\,
  F_{zv}^2\right] \,. \label{eom3}
\ea
In the right hand side of these expressions a subindex in $f$ denote
partial derivative, $f_v\equiv\partial_vf$ and
$f_z\equiv\partial_zf$.

A crucial point to be verified is whether the infalling matter that
supports the time dependent solution satisfies appropriate energy
conditions. In particular, the weakest energy condition that can be
imposed is the so called null energy condition, that states that for
any null vector $n^\mu$, the energy momentum tensor must satisfy
\be T_{\mu\nu}^{\sf Matter} n^\mu n^\nu  \geq  0\,, \label{NEC} \ee
everywhere in spacetime. Writing a generic vector as
$n^\mu=(n^v,n^z,n^x,{\bf 0}_{d-2})$, where we have used the
rotational invariance in the Cartesian coordinates ${\bf x}_{d-1}$
to eliminate redundant components, the condition for it to be null
reads
\be g_{\mu\nu}n^\mu n^\nu = \frac 1{z^2}(-f (n^v)^2 - 2n^v n^z +
(n^x)^2)=0\,. \label{null} \ee
This equation can be solved for $n^z$. There is one solution
$n^v=n^x=0$ under which \eqref{NEC} vanishes identically imposing no
constraint. On the other hand, when $n^v\neq 0$ the solution reads
$n^z = ((n^x)^2-f (n^v)^2)/2n^v$, and upon replacing it into the
energy condition \eqref{NEC} one finds
\be (T_{vv}^{\sf Matter }-fT_{vz}^{\sf Matter })(n^v)^2+(T^{\sf
Matter }_{xx}+T^{\sf Matter}_{vz})(n^x)^2 \geq 0\,. \ee
In order for this quadratic form to be positive definite for any
null vector, we need $(T_{vv}^{m }-fT_{vz}^{m })\geq 0$ and $(T^{m
}_{xx}+T^{m }_{vz})\geq 0$, using \eqref{eom}-\eqref{eom3} one finds
\ba
 &&f_v  \leq  0 \,,
 \label{cond1}\\
 &&
 z f_{zz} - (d-1)f_z - 2{z^3} \, F_{zv}^2  \geq  0 \,.
\label{cond2} \ea
This set of equations constraint the $(z,v)$-dependence of the
ansatz functions $f,F_{zv}$.

Another important requirement to be imposed on the solution is that
any physical matter current sourcing the gauge fields must be
time-like or null. Defining the matter current as
\ba
j^\nu_{\sf Matter} &=&\nabla_\mu F^{\mu\nu}\nonumber\\
 &=& \frac 1{\sqrt{-g}}\,\partial_\mu \left(\sqrt{-g}F^{\mu\nu}\right)\,,
 \label{max}
\ea
we find
\ba j^z_{\sf Matter}&=&\frac{z^4}{L^3} F_{zv,v}\,, \label{jj0}
\\
j^v_{\sf Matter}&=&\frac{({d-3})z^3}{L^3} F_{zv}-\frac{z^4}{L^3}
F_{zv,z}\,. \label{jj} \ea
The matter current must satisfy
\be g_{\mu\nu}\,j^\mu_{\sf Matter}\, j^\nu_{\sf Matter} \leq 0\,,
\label{constrcurr} \ee
or using \eqref{jj}
\be \left(\frac{({d-3})}z F_{zv}- F_{zv,z}\right)\left(2
F_{zv,v}+f\left(\frac{({d-3})}z F_{zv}- F_{zv,z}\right)\right)\geq
0\,. \label{corriente} \ee

We now consider the particular case where the  $f,F_{zv}$ functions keep the
same $z$-dependence as in \eqref{soluf}-\eqref{soluphi}, and $M$ and
$Q$ become $v$-dependent functions $\hat M(v)$ and $\hat Q(v)$. As
expected, this implies that the locus of zeros of
$f$  ({\it i.e.} the horizon positions $\hat z_\pm$) will be
$v$-dependent, and  equations \eqref{jj0}-\eqref{jj} imply that
$j^\mu_{\sf Matter}=(0,j^z_{\sf Matter} ,{\bf 0}\,)$ with $j^z_{\sf
Matter}$ given by \eqref{jj0}. This matter current satisfies
condition \eqref{constrcurr} as an equality, which means that the
charged source for the gauge field is light-like.  Moreover, the
ensuing $T_{\mu\nu}^{\sf Matter}$ takes a null dust form, and the
energy condition \eqref{cond2} is again satisfied as an equality. In
passing, we notice that although we have a time dependent electric
field, the matter current leads to no magnetic field in Ampere's
law. In summary, generalizing the ansatz to $v$-dependent functions
$\hat M(v)$ and $\hat Q(v)$ in $f$ and $F_{zv}$ leads to a
physically sensible gravity solution as long as they satisfy
\eqref{cond1}, which can be rewritten as
\be \hat M_v\geq 2\hat Q\hat Q_v z^{d-2} \,. \label{reduced} \ee
This energy condition \eqref{reduced} was discussed in
\cite{martin,subad}. We would like to remark that \eqref{reduced}
coincides with the condition for the horizon not the recede. Indeed,
if we define the (time dependent) horizon position $\hat z_h$ as the
solution to $f(\hat z_h,v)=0$, we get $d\hat z_h/dv=-f_v/f_z \propto-\hat M_v+2\hat
Q\hat Q_v z^{d-2}$, thus from the condition $d\hat z_h/dv \leq 0$ we
obtain \eqref{reduced} (see Fig. \ref{japi} below). Nevertheless, condition \eqref{reduced} must
be satisfied even in the absence of horizons.

We now turn to analyze its consequences:
\begin{itemize}
\item
When the charge $\hat Q$ vanishes,  \eqref{reduced} translates into
\ba \hat M_v\geq 0\,. \label{condd} \ea
Therefore, any interpolation between pure AdS and the planar
AdS-Schwarzschild metric with a monotonically growing mass function,
asymptotically reaching a constant value $M$,  is a healthy solution
of Einstein-Maxwell equations. The required additional matter
contribution satisfies the null energy condition. The chargeless
Vaidya metric case, used in \cite{bala} to study thermalization of a
strongly coupled plasma at zero chemical potential after an energy
quench, allows for an arbitrary time pattern of energy injection
(this is a well known result, and we mention it here only for
completeness).
\item If the charge remains constant as we perform the quench ($\hat Q_v=0$),
condition \eqref{reduced} reduces to \eqref{condd}. As discussed
above, the constraint \eqref{reduced} is identically satisfied as
long as the mass function is monotonically increasing. The
corresponding Vaidya geometry can be used to study thermalization
processes at fixed chemical potential   after an injection of
energy.
\item For non-constant charge situations,  outside of the support of $\hat Q_v$,
condition \eqref{reduced} reduces to \eqref{condd}. On the other
hand, for analytic $\hat Q$ the support is non-compact, implying
 \begin{itemize}
 \item
Whenever $\hat Q\hat Q_v>0$, it is immediate to see that
\eqref{reduced} is violated at large enough $z$ for any fixed value
of $v$. Notice that $\hat Q\hat Q_v>0$ corresponds to an increasing
absolute value of the charge, what we will call a ``charging''
background in what follows. The null energy condition therefore tell
us that we can use a charging AdS Vaidya ansatz to study the
thermalization of the dual field theory, as long as  our probes
explore the near boundary region. From the dual point of view this
means that the geometry can only be trusted to describe the UV
degrees of freedom of the field theory. These classes of solutions
were used in \cite{martin, subad} to study the thermalization
process after a quench on energy and chemical potential.
\item No additional constraint follows from \eqref{reduced} for the case of a
``discharging'' background $\hat Q\hat Q_v<0$, provided the mass is
increasing $M_v>0$.
\end{itemize}
\end{itemize}

In summary,  the charged metrics obtained by promoting the AdSRN
charge and the mass into $v$-dependent functions (with growing mass), satisfy energy
conditions everywhere in spacetime whenever they are ``discharging''
$\hat Q\hat Q_v<0$, while violate energy conditions in the bulk for
large enough $z$, {\em i.e.} in the deep IR, whenever they are
``charging'' $\hat Q\hat Q_v>0$. In this last case probing the near
boundary geometry is still a safe option. Note that in the above
definitions, and throughout this paper, we use the words
``charging'' and ``discharging'' as referring to an increasing or
decreasing {\em absolute value} of the charge.

In what follows, we use the aforementioned charging and discharging
geometries in order to get information about the thermalization
processes of the boundary theory. To such end, we choose our
functions $\hat M$ and $\hat Q$ as interpolating between initial
constant values $M_{\sf in}$ and $Q_{\sf in}$ in the asymptotic past
$v\to -\infty$ and final constant values $M$ and $Q$ in the
asymptotic future $v\to\infty$.

In the time-dependent background, the field strength in
\eqref{RNADSBHA} can be obtained from the gauge potential
\be A_v= -\gamma  \hat QL \,z^{d-2} +  \hat \mu L\,, \label{amu} \ee
with $\hat\mu$ an arbitrary $v$-dependent function. The background
$v$-dependence precludes the global definition of a time coordinate,
since \eqref{time} is not an exact differential, avoiding an
Euclidean continuation. Thus, the only constraint we impose on the
function $\hat \mu$ arises from the regularity of the Euclidean
continuation of the asymptotic regions $v\to\pm\infty$.  In other
words we choose $\hat\mu\to \gamma Q_{\sf in} z_{\sf in}^{d-2}$ when
$v\to- \infty$, and $\hat\mu\to\gamma Q z_h^{d-2}$ when
$v\to\infty$, where $z_{\sf in}$ and $z_{h}$ are the positions of
the horizons before and after the quench.
%
%
%
%
%
%
\subsubsection*{Boundary theory}
\label{time-boundary}
%
%
%
%
In the dual field theory, the proposed geometry describes a system
interpolating between two different values of the temperature as the
result of injecting energy homogeneously into the system, and at the
same time quenching the chemical potential. The quench can take the
chemical potential up or down depending on whether we have a
charging or a discharging geometry, and corresponds to a homogeneous
injection of particles/antiparticles into the system. Even if, as
mentioned in the previous sections, there is no global time
coordinate, with the hindsight of the static background we define a
time coordinate in the asymptotic region $z\to0$ as $t\sim v$: the
variable $v$ at the boundary then coincides with the gauge theory
time.
%
%
%
%
%
%
\section{Probes of thermalization}
\label{probes}
%
%
As a probe for analyzing the  thermalization process, we study equal
time two point correlators of charged scalar operators
$\mathcal{O}_\Delta$ of conformal dimension $\Delta$. Decomposing
the bulk coordinates as ${x}^\mu =(x^\alpha,z)$, the AdS/CFT
dictionary relates the QFT two point correlator
$\langle\mathcal{O}_{\Delta}(x^\alpha_1)\mathcal{O}_{\Delta}(x^\alpha_2)\rangle$
to the Feynman propagator $G({x}_1^\alpha,z_1 |\,{x}^\alpha_2,z_2 )$
of a charged scalar field of mass $m=\sqrt{\Delta(\Delta-d)}/L$
propagating in the bulk, according to the formula \cite{banks}
\be
\langle\mathcal{O}_{\Delta}(x^\alpha_1)\mathcal{O}_{\Delta}(x^\alpha_2)\rangle
= \lim_{z_1,z_2\to
0}\,z_1^{-\Delta}{z_2}^{-\Delta}G(x^\alpha_1,z_1|\,x^\alpha_2,z_2)\,.
\label{correlator}
\ee
In the large mass limit $1\ll mL\approx\Delta $, the bulk propagator
can be approximated by  the classical trajectory (see appendix
\ref{Worldline})
\ba
&&G(x^\alpha_1,z_1|\,x^\alpha_2,z_2)\approx e^{i S_{\sf
on-shell} (x^\alpha_1,z_1|\,x^\alpha_2,z_2)}\,,\label{pin}
\ea
with
\ba
S_{\sf on-shell}(x^\alpha_1,z_1|\,x^\alpha_2,z_2)=\left. \int
d\tau \left(-m\sqrt{-g_{\mu\nu}\dot x^\mu \dot x^\nu} +  {\sf e}
A_\mu\dot x^\mu\right) \right|_{x^\mu(\tau)=x^\mu_{\sf
classical}(\tau)}\,. \label{action}
\ea
Here $S_{\sf on-shell}$ is the appropriate action for the time-like
trajectory, for a particle of mass $m$ and charge $\sf e$, evaluated
on the classical trajectory $x^\mu_{\sf classical}(\tau)$ that joins
$(x_1^\alpha,z_1)$ and $(x_2^\alpha,z_2)$. The charge to mass
quotient ${\sf e}/m$ is supposed to remain finite in the large $m$
limit (see Appendix \ref{Worldline} for details). In the limit of
large conformal dimension ${\Delta}\gg1$, equations
\eqref{correlator}-\eqref{action} instruct us to compute the
two-point correlator (\ref{correlator}) from the classical
trajectory of a charged particle whose endpoints sit at the
boundary. It is important to realize that any such trajectory lies
completely in the classically forbidden region, a complete
discussion of the different ways to access such region with
classical trajectories is given in Appendix \ref{Euclid}. For the
time being, let us analytically continue the parameter $\tau$ and
the electric charge $\sf e$ according to $\tau=-i\tau_E$ and ${\sf
e}=i {\sf e}_E$, resulting in
\ba &&G(x^\alpha_1,z_1|\,x^\alpha_2,z_2)=e^{- S^E_{\sf on-shell}
(x^\alpha_1,z_1|\,x^\alpha_2,z_2)}\,, \label{pindon} \ea with \ba
S^E_{\sf on-shell} (x^\alpha_1,z_1|\,x^\alpha_2,z_2)= m\left. \int
d\tau_E\left(\sqrt{g_{\mu\nu}{x'}^\mu {x'}^\nu} +  q_E A_\mu
x'^\mu\right) \right|_{x^\mu(\tau_E)=x^\mu_{\sf
classical}(\tau_E)}\,. \label{actionE} \ea
where we defined $q_E={\sf e}_E/m$.

In what follows we evaluate the two point correlator according to
\ba &&\langle\mathcal{O}_{\Delta}( x^\alpha_1)\mathcal{O}_{\Delta}(
x^\alpha_2)\rangle =\lim_{z_\epsilon\to 0}z_\epsilon^{-2\Delta}
e^{-S^E_{\sf on-shell}(x^\alpha_1,z_\epsilon|\,x^
\alpha_2,z_\epsilon)}\,, \label{donga} \ea
where $z_\epsilon$ is a cutoff in holographic direction, and the
on-shell Euclidean action $S^E_{\sf on-shell}(x^
\alpha_1,z_\epsilon;x^\alpha_2,z_\epsilon)$ is given in
(\ref{pindon}), and evaluated on classical paths starting at
$(x^\alpha_1,z_\epsilon)$ and ending at $(x^\alpha_2,z_\epsilon)$.
In AdS space any geodesic approaching the boundary has a divergent
logarithmic contribution to the length $\sim -\log z$, the
$z^\Delta$ factors in \eqref{correlator} are present to precisely
cancel this contribution. In Appendix \ref{WKB} we show that
expression \eqref{donga} for the correlator coincides with the
standard holographic definition, as the quotient of the subleading
to the leading components of the bulk scalar field as $z$ approaches
the boundary, when the Klein-Gordon equation is solved in the WKB
approximation.

Therefore, to evaluate the two-point correlator for the quench we
insert in (\ref{actionE}) the time dependent background of section
\ref{timedep}. Afterwards, we compare it with the corresponding
correlator obtained by substituting in formula (\ref{donga}) the
static background of section \ref{thermal}. The probed degrees of
freedom can be said to have reached thermal equilibrium, whenever
the quantity
\ba
\delta S = S^E_{\substack{\!\!\!\!\!\!\!\!\!\!\sf
on-shell\\ \sf t -dependent}} - S^E_{\substack{\!\!\!\!\sf
on-shell\\\sf equilibrium}}\,,
\label{thermalz}
\ea
vanishes. Our interest is to find a time profile of $\delta S$.

\begin{figure}[h]
\bc \vspace{-0.3cm}
\includegraphics[width=14cm]{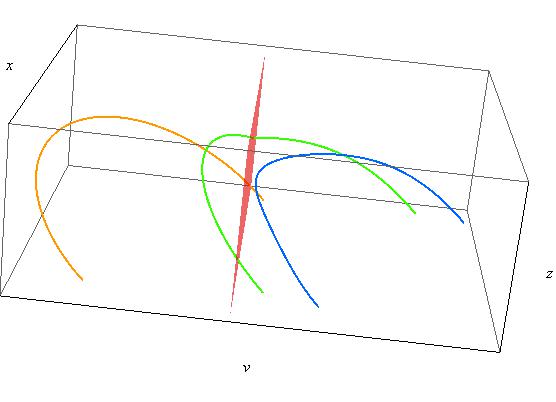}
\vspace{-0.5cm} \caption{{\sf U}-shaped profiles of the geodesics
required to compute the two-point correlators. Gauge theory time
runs horizontally to the right and is parametrized by the
$v$-coordinate in eq. \eqref{RNADSBH}. The  endpoints of the
trajectories,  located at the AdS boundary, are separated in space
and time (see \eqref{ccculo}). The geometry to the left of the shell
is pure  AdS ($\sf T=0$), at $v=0$ energy and charge are injected
homogeneously into the system giving rise  to the formation of a
charged black hole geometry ($\sf T>0$) to the right of the shell.
Early geodesics (orange) intersect the shell at $v=0$ (red plane)
only once, late geodesics (green) intersect it twice, while
thermalized geodesics (blue) do not intersect it at all. }
\label{lapizjapones2} \ec
\end{figure}
\begin{figure}[h]
\bc
\includegraphics[width=5.5cm]{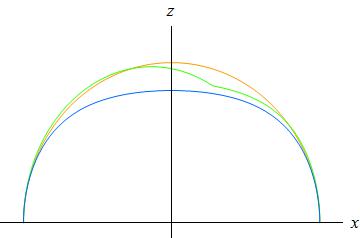}\quad
\includegraphics[width=5.5cm]{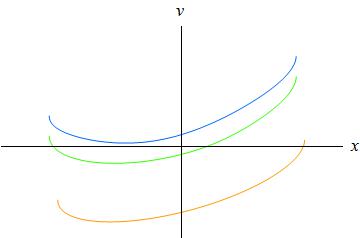}\quad
\includegraphics[width=5.5cm]{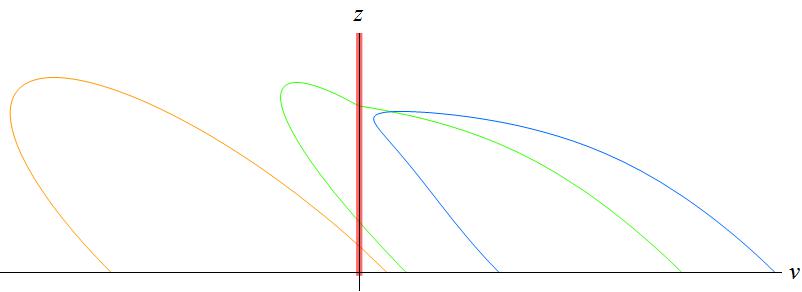}
\caption{Plots of the three types of geodesics that we encounter in
our calculations. Early geodesics (orange) intersect the shell at
$v=0$ (red line) only once, late geodesics (green) intersect it
twice, while thermalized geodesics (blue) do not intersect it at
all. The plots show the $xz$  (left), the $xv$  (center) and  $vz$
(right) projections.}
\label{lapizjapones}
\ec
\end{figure}

In the present work we look for spacelike $\sf U$-shaped
trajectories for mass $m$ and charge $\sf e$ on the the
aforementioned backgrounds starting and ending at the boundary,
whose endpoints are separated along one of the Cartesian coordinates
by a distance $\ell $, and in time by a distance $\Delta t $ (see
Figs.\,\ref{lapizjapones2} and \ref{lapizjapones}).

~~~~~~~~~~~~~~~~~~~~~~~~~~~~~~~~~~~~~

We choose to
parametrize the curve as $x^{\mu}(x)=(v(x),
z(x),x_1(x)\!=\!x,{\bf x}_{d-2})$ with ${\bf x}_{d-2}$ constant, and
without any loss of generality we take $x_i<x<x_f$, with $x_i,x_f$
the extremes of the $\sf U$-shaped curve whose tip sits at $x=0$.
The boundary conditions for the trajectories we are interested in
are
\be z(x_i)=z(x_f)=z_\epsilon,\quad \quad v(x_i)=t_i\,, \quad \quad
v(x_f)=t_f\,. \label{boundcond2} \ee
where, as mentioned above, we have regularized the problem by
considering geodesics whose endpoints lie at $z=z_\epsilon$, instead
of $z=0$ ($z_\epsilon\to0$ should be taken at the end of the
computations). Alternatively, the trajectories can be characterized
in terms of the boundary conditions at the tip
\be
z(0)=z_*,\quad \quad v(0)=v_*\,,\quad \quad v'(0)=v'_*\,,\quad \quad
\quad z'(0)=0\,.
 \label{tip}
\ee
Viewed from the tip, two branches appear: (i) for $x_i<x<0$ the
trajectory describes $z\in(0,z_*)$  and hence $z'>0$, (ii) while for
$0<x<x_f$ it goes from the tip $z=z_*>0$ to the boundary $z=0$
therefore $z'<0$. These two branches correspond to the two signs in
\eqref{zprima}

\begin{figure}[h]
\bc
\includegraphics[width=7.5cm]{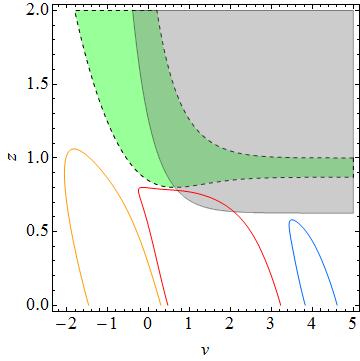}
\caption{Geodesics and the null energy condition: the grey zone
depicts the (IR) region in which the null energy condition
\eqref{reduced} is violated.  We also depict the position
of the outer/inner horizon, defined as the $z(v)$ solution to
$f(z,v)=0$, as dashed lines. The yellow and blue geodesics do not explore
the sick region and can be used to analyze thermalization. The
red geodesics on the other hand enter into the sick region and we
cannot trust them to probe the thermalization process, in other words the
background geometry can be used to analyze the thermalization of
degrees of freedom above an IR cutoff. Note that in concordance with the discussion
below \eqref{reduced}, the region where the horizon recedes does not satisfy the energy
conditions. The background interpolates between pure $AdS$ in
the past to a $Q=2$ and $M=1+Q^2$ black hole in the future.} \label{japi} \ec
\end{figure}

The space and time separation characterizing the geodesic  are given
by
\be
\ell = x_f-x_i\,, \qquad \qquad\qquad \Delta t= t_f-t_i\,.
\label{ccculo}
\ee
At this point we can already suspect that $\ell$ will be a growing
function of $z_*$, implying that any upper cutoff in $z_*$ imposes
an upper cutoff in $\ell$. This behavior is well known and can be
explicitly checked in the numerical calculations of the forthcoming
sections. Moreover, this implies that if we are interested in
restricting the region of the geometry to be probed by our geodesics
to the near boundary patch, then we must impose an upper bound on
$\ell<\ell_{\sf Max}$. In other words, the violation of energy conditions in
the bulk for the charging case imposes an IR cutoff in the degrees
of freedom that can be probed (see fig.\,\ref{japi}).

~

%
%
%
%
%
\subsection{Equilibrium state}
\label{probe-eq}
%
%
%
Inserting the eternal AdSRN solution \eqref{RNADSBH-st}  into the
action \eqref{actionE} we get
\ba S^E_{\sf equilibrium} &=&mL\int_{x_i}^{x_f}dx\left(
\frac1z\sqrt{1-f\,v'^2-2z'v'}-q_E(\gamma{ Q}z^{d-2}-\mu)v' \right)
\,, \label{act} \ea
where prime $(')$ denotes derivative with respect to the Euclidean
parameter $\tau_E$, that we have gauge fixed to $\tau_E=x$. The
first term corresponds to the geodesic length, while the second
codifies the coupling to the gauge potential. Notice that only for
the case of geodesics starting and ending at the same value of $v$
can one drop the $\mu$ contribution on the second term.

The two resulting second order equations of motion can be shown to
have two first integrals: one coming from reparametrization
invariance, which in our coordinates turns into the conserved
Hamiltonian generating translation along the $x$ parameter, and the
second from $v$-independence of the metric. The equations to be
solved are\footnote{For the case of a chargeless particle,
eqn \eqref{vinc} implies that the particle motion can be constrained
to the $t=t_0$ surface.}
\ba z\sqrt{1-f\,v'^2-2z'v'}&=&\frac 1 {\tilde E} \,, \label{viinc}
\\
(f\,v'+z')\tilde E+\gamma q_EQz^{d-2}&=&f_*\,v'_* \tilde E+\gamma
q_EQ z_*^{d-2}\,, \label{vinc} \ea
where the value of the constants on the right hand side have been
fixed at the tip of the $\sf U$-shaped curve ($x=0$) according to
the conditions \eqref{tip}: ${\tilde E}=1/z_*\sqrt{1-f_*\,v_*'^2}$
where $f(z_*)=f_*$.

Solving \eqref{viinc}-\eqref{vinc} for $z'$ one obtains
\be
z'= \pm\sqrt{\left( \frac{1}{\tilde E^2 z^2}
-1\right)f+\left(f_*v'_*+\frac{\gamma q_EQ}{\tilde E}
(z_*^{d-2}-z^{d-2})\right)^2}\,. \label{zprima}
\ee
As explained above, the double sign in this expression corresponds
to the two branches of the {\sf U}-shaped trajectory:  the positive
sign corresponds to $x_i<x<0$ while the negative one happens for
$0<x<x_f$. Solving for $v'$ one obtains
\be
v'=\frac1f \left( \frac{\gamma q_EQ}{\tilde
E}(z_*^{d-2}-z^{d-2})+f_*\,v'_* -z' \right)\,,
\label{vprimastat}
\ee
with $z'$ given by \eqref{zprima}. The two parameters at the tip
($z_*,v'_*$) are related to ($\ell,\Delta t$) by noting that
\be \ell = x_f-x_i = \int_{x_i} ^{0}dx+\int_0
^{x_f}dx=2\int_{0}^{z_*}\frac{dz}{z'}
\,, \ee
and
\be
\Delta t = t_f-t_i =v(x_f)-v(x_i)=\int_{x_i}^{x_f} dx \,v'\,.
\ee
Finally, inserting \eqref{viinc} and \eqref{vprimastat} into
\eqref{act} we can express the on-shell action as\footnote{Since the
trajectories endpoints locate at the same radial position, a total
derivative term vanishes.}
\be
S^E_{\substack{\!\!\!\!\sf on-shell\\\sf equilibrium}}= 2\frac{
mL} {\tilde E} \int_{z_\epsilon}^{z_*} \frac{dz}{z'}\left(
\frac{1}{z^2}+ \gamma q_E Q \frac {z^{d-2}}f \left(\gamma q_E
Q(z^{d-2}-z_*^{d-2})- f_*v'_*\tilde E\right) \right) +mL\,\mu
q_E\Delta t\,.
\label{act-os}
\ee
Since the on-shell action diverges when the endpoints of the
trajectory reach the boundary, we have regularized \eqref{act-os}
by introducing a cutoff $z_\epsilon$. The factor of two arises from
the two branches of the trajectory giving identical contributions.
The formulae above enable us to compute the on shell action
\eqref{act-os} as a function of $\ell$ and $\Delta t$. Notice that
the $v$-independence of the background implies that no dependence on
$v_*$ is expected.
%
%
%
%
%
%
\subsection{Time dependent state}
\label{probe-timedep}


%
The action for a charged particle moving in the time dependent
metric takes the form \eqref{act}, but with the constants $Q$ and
$M$ substituted by the $v$-dependent functions $\hat Q$ and $\hat
M$.
\ba S^E_{\sf t-dependent}&=&mL\int_{x_i}^{x_f}dx\left(
\frac1z\sqrt{1-f\,v'^2-2z'v'}-q_E(\gamma{ \hat Q}z^{d-2}-\hat \mu)v'
\right) \,. \label{act-time} \ea
Notice that we have also introduced a $v$-dependent $\hat \mu$ (see
discussion below \eqref{amu}). Due to the lack of $v$-translation
invariance, the on shell action cannot be taken into a pure $z$
integral as in \eqref{act-os}. The equations to be solved now are
%
\ba z\sqrt{1-f\,v'^2-2z'v'}&=&\frac{1}E\,, \label{viinc-tb}
\\
E\left(f v''+z''+\frac{f_v v'^2}{2}+f_z z' v'\right) &=&-
\gamma{q_E}\hat Q(d-2)z^{d-3} z' \,, \label{eom-td} \ea
%
where the value of $E$ is again fixed at the tip of the curve \eqref{tip} and given by
${  E}=1/z_*\sqrt{1-f_*\,v_*'^2}$.

These equations are solved numerically by the shooting method. In
practice, we shoot from the turning point $z_*$ with initial (final)
conditions (\ref{tip}) for $0<x<x_f$ and $x_i<x<0$. For each choice of
values $(z_*,v_*,v'_*)$ at the tip we will find an $\sf U$-shaped
geodesic characterized by $(\ell,\Delta t,t_f)$ (see
Figs.\,\ref{lapizjapones2} and \ref{lapizjapones}). The plots below
give the thermalization curves for $\delta S$ given by \eqref{thermalz} with the on-shell action
\eqref{act-time} being
generically a function $S^E_{\substack{\!\!\!\!\!\!\!\!\!\!\sf
on-shell\\ \sf t-dependent}}\!\!\!\!\!\!(\ell,\Delta t,t_f)$.

~

Notice that, since the time dependent charging background violates
the null energy conditions for large enough $z$, in such case we
need to make sure that our geodesics are only probing the healthy
part of the geometry (see Fig. \ref{japi}). As mentioned before, this implies that  we can
probe thermalization with correlation functions with bounded
$\ell<\ell_{\sf Max}$. On the other hand in the discharging background, we
have no such IR cutoff and we can probe the thermalization with any
value of $\ell$.
%
%
%
%
%
%
%
\section{Results}
\label{results}
%
%
%
%
%
The considerations of the previous sections were made for arbitrary
functions $\hat M$ and $\hat Q$. In this section, in order to
proceed with the numerical evaluation of the thermalization time for
different probes, we will need explicit expressions for those
functions. We choose
\ba \hat M&=&\frac{M-M_{\sf
in}}{2}\left(1+\tanh\left(\frac{v}{v_0}\right)\right)+M_{\sf in}\,,
\label{acapapa1}
\\
\hat Q&=&\frac{Q-Q_{\sf
in}}{2}\left(1+\tanh\left(\frac{v}{v_0}\right)\right)+Q_{\sf in}\,.
\label{acapapa2} \ea
Here $v_0$ parametrizes the shell thickness and the $v_0\to 0$ case
corresponds to the shock wave discussed in \cite{bala}. These
functions satisfy
\ba && {\lim_{v\to {-\infty} }}\hat M = M_{\sf in}\,, \qquad\qquad
\qquad {\lim_{v\to \infty }}\hat M = M \n && {\lim_{v\to{-\infty}
}}\hat Q = Q_{\sf in}\,, \qquad\qquad \qquad\;\; {\lim_{v\to \infty
}}\hat Q = Q\,. \ea
Therefore the backgrounds we consider interpolate between an AdSRN black hole
with mass $M_{\sf in}$ and charge $Q_{\sf in}$ in the distant past
$v\ll v_0$ and an AdSRN black hole with mass $M$ and charge $Q$ in
the distant future $v \gg v_0$.

%
%
%
%
%

\subsection{Vanishing background charge}

We start by analyzing the thermalization process in the case of vanishing
background charge, which corresponds to a thermal quench with vanishing chemical potential
in the boundary theory. To this end, we choose
\ba \hat M&=&\frac{M-M_{\sf
in}}{2}\left(1+\tanh\left(\frac{v}{v_0}\right)\right)+M_{\sf in}\,,
\\
\hat Q&=&0\,. \ea
For the case of vanishing $\Delta t$ we reproduced the results of
\cite{bala}, and for nonvanishing $\Delta t$ we find results in agreement
 to those in \cite{Aparicio:2011zy}.

\begin{figure}[h]
\bc
\includegraphics[width=5.6cm]{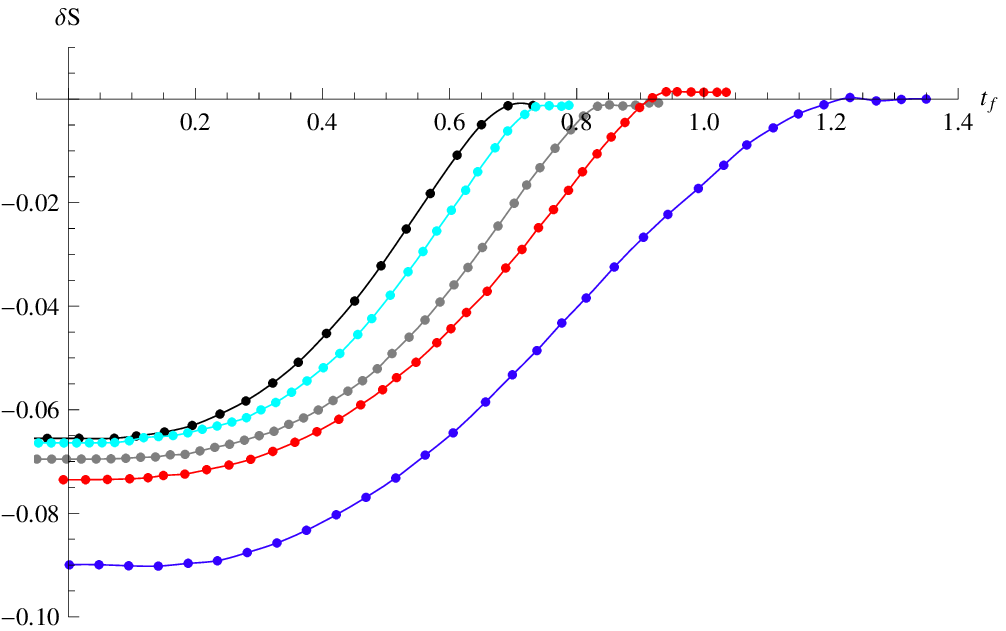}
\includegraphics[width=5.6cm]{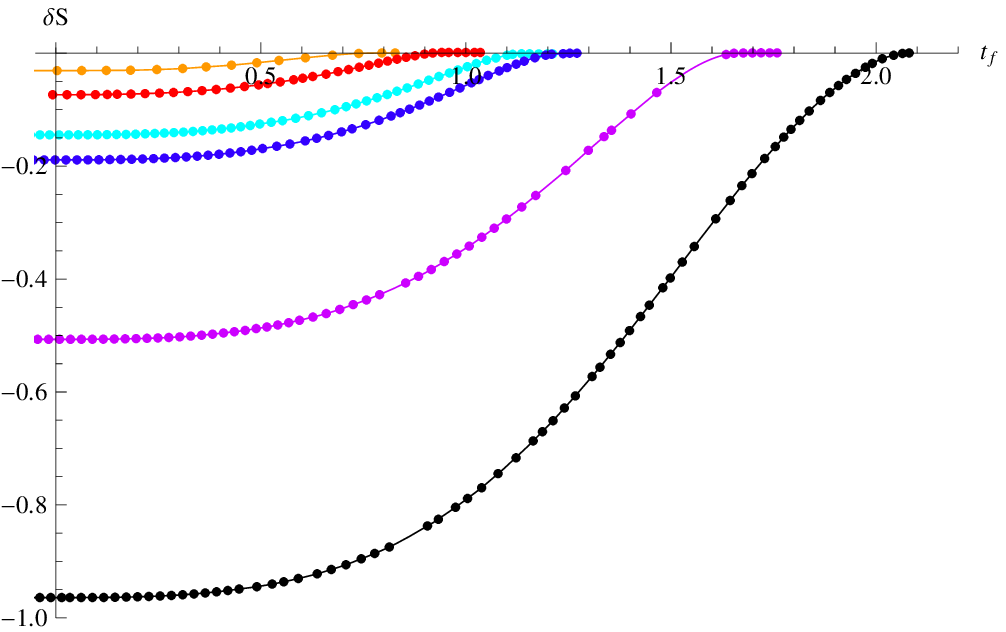}
\includegraphics[width=5.6cm]{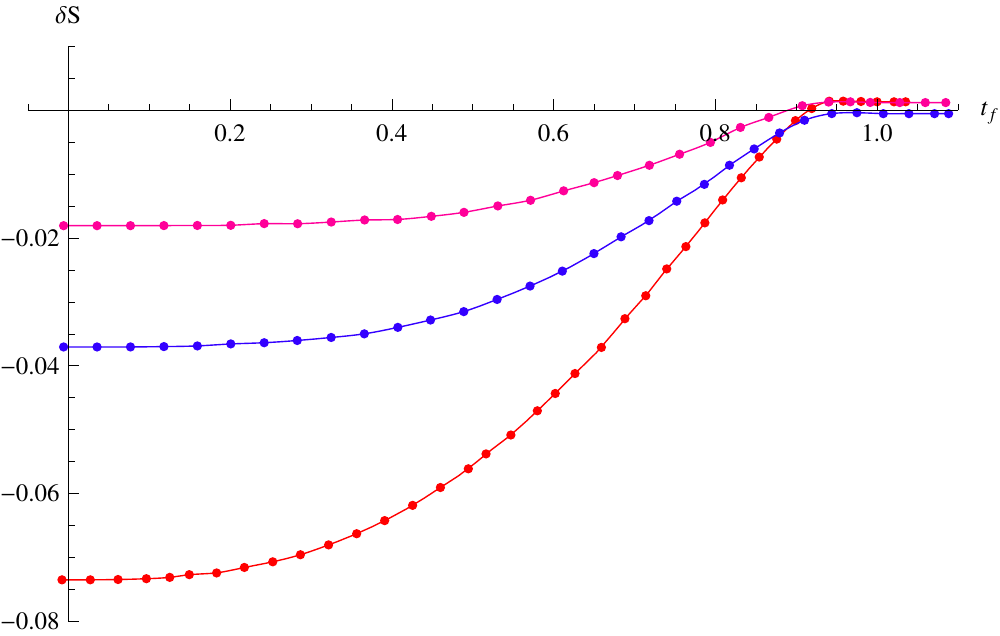}
\caption{Two point correlators as a function of
$t_f$ for a thermal quench  in the absence of chemical potential (Sect IV.A). 
~~{\it Left}: plots for $d=3$, $\ell =1.4$ and
$\Delta t=0,0.1,0.3,0.5,1$ (black, azure,
gray, red and blue  resp.). 
~~{\it Center}: plot for $d=3$, $\Delta t=0.5$ and
$\ell=1 ,1.4,1.8,2,3,4$ (orange, red, azure, blue, violet and black
resp.).
~~{\it Right}: plot for $\ell=1.4$, $\Delta t=0.5$ and $d=3,4,5$ (red, blue and
violet resp.). 
All cases correspond to a chargeless background interpolating between $M_{\sf in}=0$
and $M=1$.}
\label{balal}
\ec
\end{figure}

The results are summarized in Fig. \ref{balal}, where the
thermalization curves $\delta S$ involving the on-shell action are plotted as functions
of $t_f$. The thermalization time, defined as the approximated value of
$t_f$ at which the curve reaches the horizontal axis $\delta S=0$,
increases with $\ell$ and with $\Delta t$, implying that UV degrees
of freedom thermalize first. This is the phenomenon known as ``top
down thermalization''. Moreover, we see that the thermalization time
also increases with the dimension of the system.
\vspace{-1cm}

\subsection{Vanishing probe charge}

We now consider a quench in both the temperature and the chemical potential.
We first study the effects of the background on a vanishing probe
charge $q_E=0$. For the sake of illustration, we set our
background to be pure AdS (${\sf T}=\mu=0$) in the asymptotic past, and AdSRN (${\sf T},\mu\ne0$) in the
asymptotic future, namely
\ba \hat
M&=&\frac{M}{2}\left(1+\tanh\left(\frac{v}{v_0}\right)\right)\,,
\\
\hat
Q&=&\frac{Q}{2}\left(1+\tanh\left(\frac{v}{v_0}\right)\right)\,. \ea

\begin{figure}[ht]
\bc
\includegraphics[width=7.5cm]{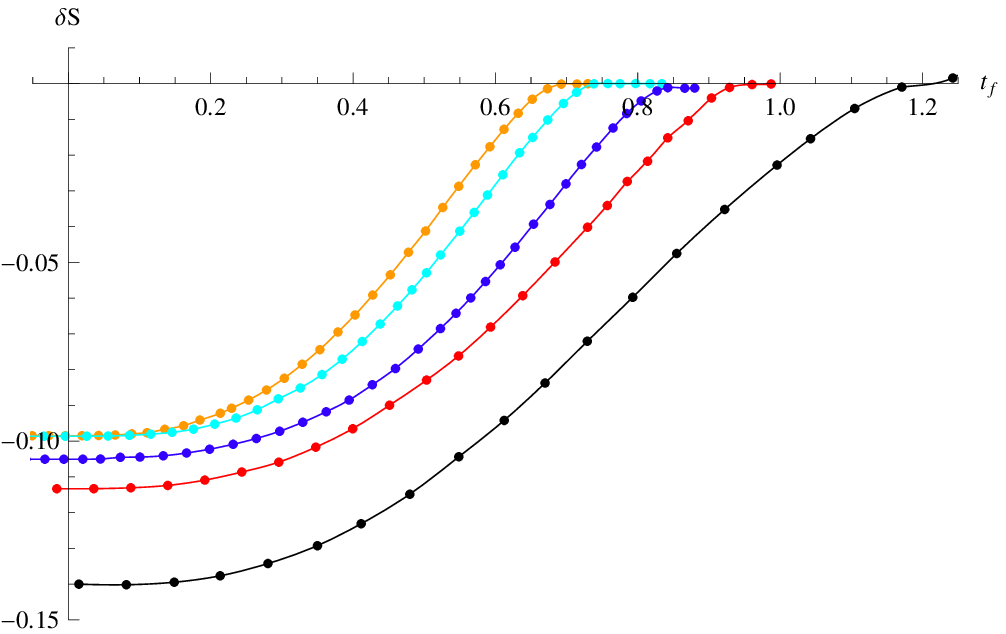}
\includegraphics[width=7.5cm]{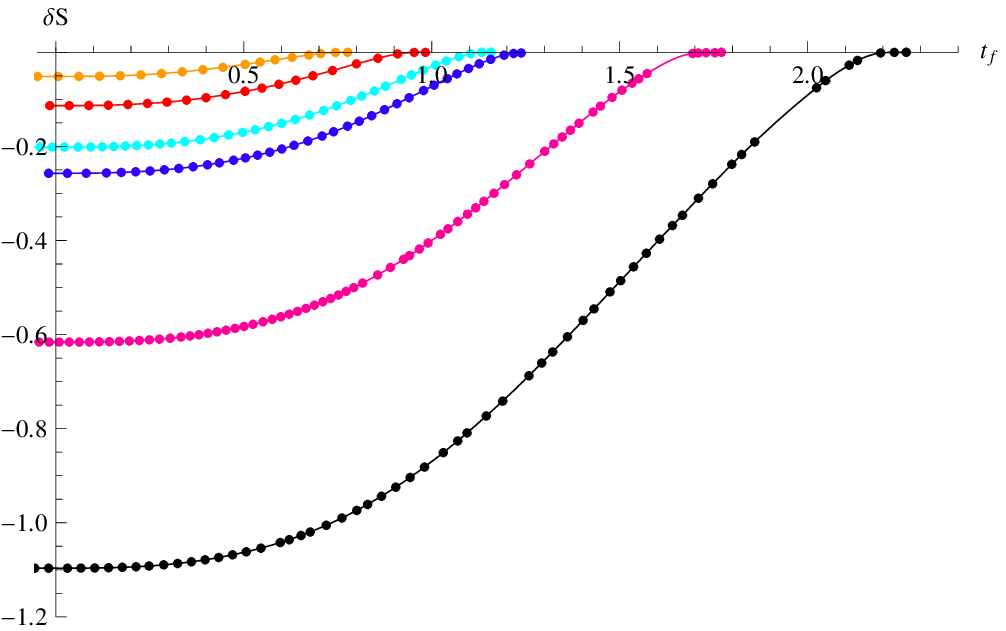}\\
\includegraphics[width=7.5cm]{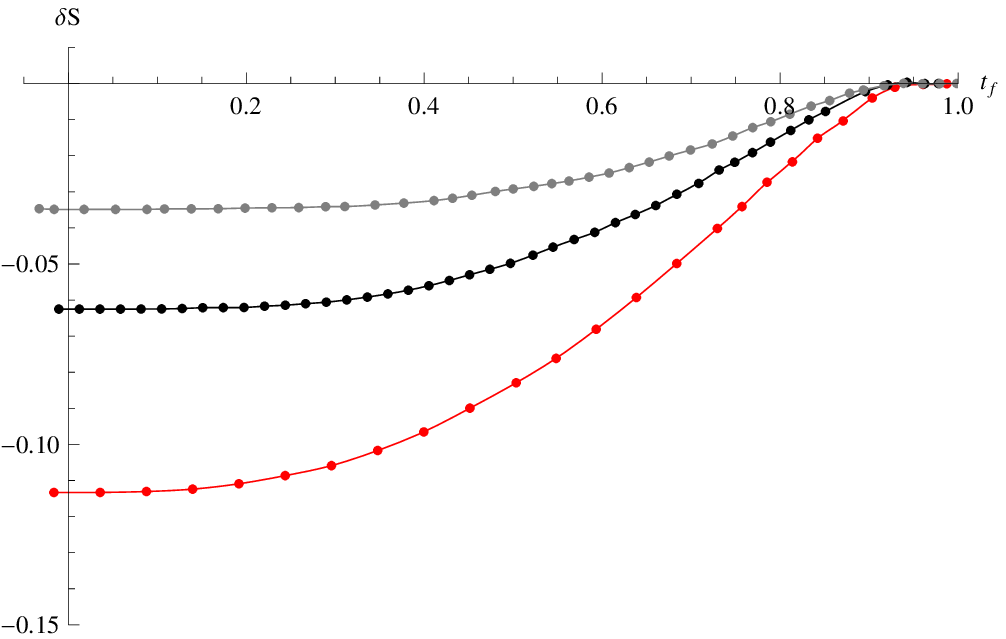}
\includegraphics[width=7.5cm]{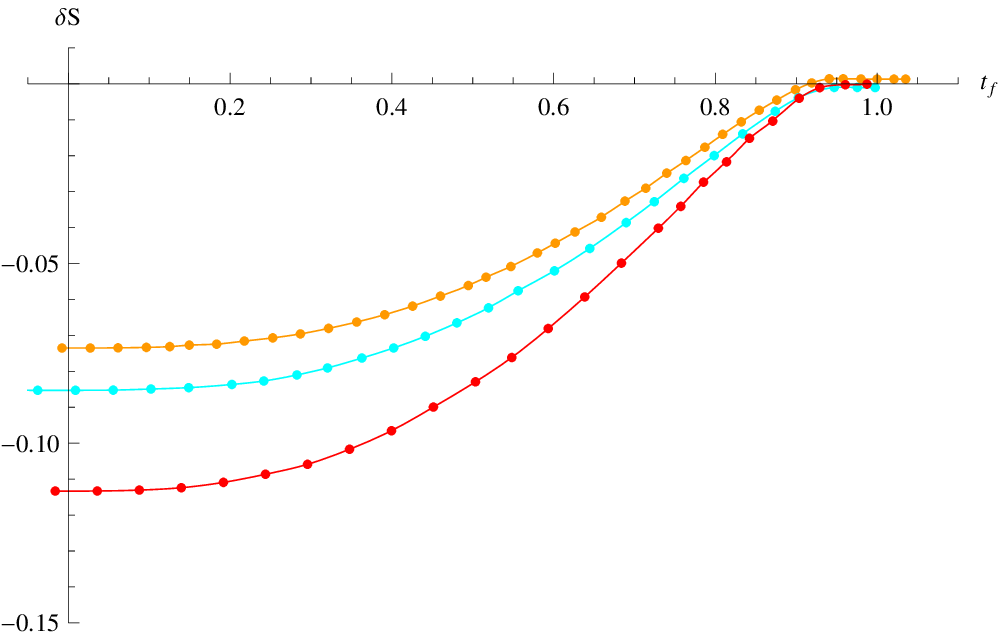}
\caption{Two point functions for uncharged operators as a function of
$t_f$ for a an energy and charge quench (Sect IV.B).
{\it Top left}: plots for $d=3$, $\ell
=1.4$  and $\Delta t=0,0.1,0.3,0.5,1$ (orange, azure,blue,
red and black  resp.).
{\it Top right}: plots for
$d=3$, $\Delta t=0.5$, and $\ell=1 ,1.4,1.8,2,3,4$ (orange, red, azure, blue, violet and black
resp.).
{\it Bottom left}: plots for
$\ell=1.4$, $\Delta t=0.5$ and $d=3,4,5$ (red, black and grey resp.). 
{\it Bottom right}: plots for $d=3$, $\ell=1.4$, $\Delta t=0.5$ and $Q=0,0.5,1$ (orange, azure and  red  resp.).
In all cases we start from pure AdS at past infinity to $M=1$ at future infinity, for the first three cases the final charge is $Q=1$.} \label{martl} \ec
\end{figure}

As could have been expected, since we are probing the system with
uncharged operators ($q_E=0$), the background charge $\hat Q$ has little effect
in the form of the thermalization curves. The phenomenon of top-down
thermalization is again present, and the thermalization time grows with
the dimension of the field theory as in the previous case. The results are depicted in Fig.
\ref{martl} and are in agreement with those of Refs. \cite{martin}.

\subsection{Constant background charge}
\label{constant}
%
%
To analyze the thermalization process for a thermal and chemical potential quench we
consider the $\hat M$ and $\hat Q$ functions to be
\ba \hat M&=&\frac{M-M_{\sf
in}}{2}\left(1+\tanh\left(\frac{v}{v_0}\right)\right)+M_{\sf in}\,,
\\
\hat Q&=&Q_{\sf in}\,. \ea
This background interpolates between a AdSRN black hole
of mass $M_{\sf in}$ and charge $Q_{\sf in}$ in the distant past
$v\ll v_0$, to a heavier AdSRN black hole ($M>M_{\sf in}$) in the distant future $v\gg
v_0$  preserving the charge $Q_{\sf in}$. In the following
we choose to  re-scale the $z$-coordinate  so as to have
$M_{\sf in}=1+Q_{\sf in}^2$ in the far past. This ensures that the background satisfies
the bound (\ref{hori}) in the past, and since the evolution increases the mass
while keeping the charge constant, \eqref{reduced} is also satisfied. Notice that
although the background charge is constant, the chemical potential changes when injecting
energy into the system. The reason for this is that we should demand regularity of the
euclidean rotated asymptotic geometries (see discussion after eq. \eqref{amu}).
We choose the profile for the chemical potential to be
\ba
\hat\mu=\frac{\mu-\mu_{\sf in}}{2}\left(1+\tanh\left(\frac{v}{v_0}\right)\right)+\mu_{\sf
in}\,,
\ea
with
\ba
\mu_{\sf in}=\gamma Q_{{\sf
in}},\quad\quad\mu=\gamma  Q_{{\sf in}} z_h^{d-2}\,,
\ea
where $z_h<1$
is the horizon position after the quench. The initial position being
$z_{\sf in}=1$ due to the condition $M_{\sf in}=1+Q_{\sf in}^2$.  Notice
that during the quench the absolute value of the chemical potential reduces, but it cannot reach $\mu=0$.

\begin{figure}[ht]
\bc
\includegraphics[width=7.5cm]{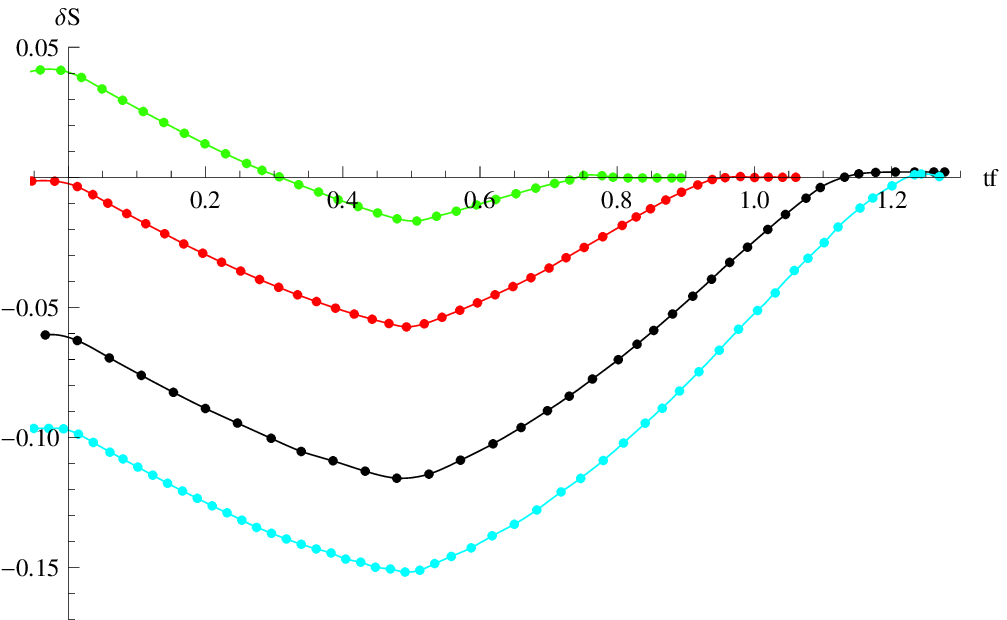}
\includegraphics[width=7.5cm]{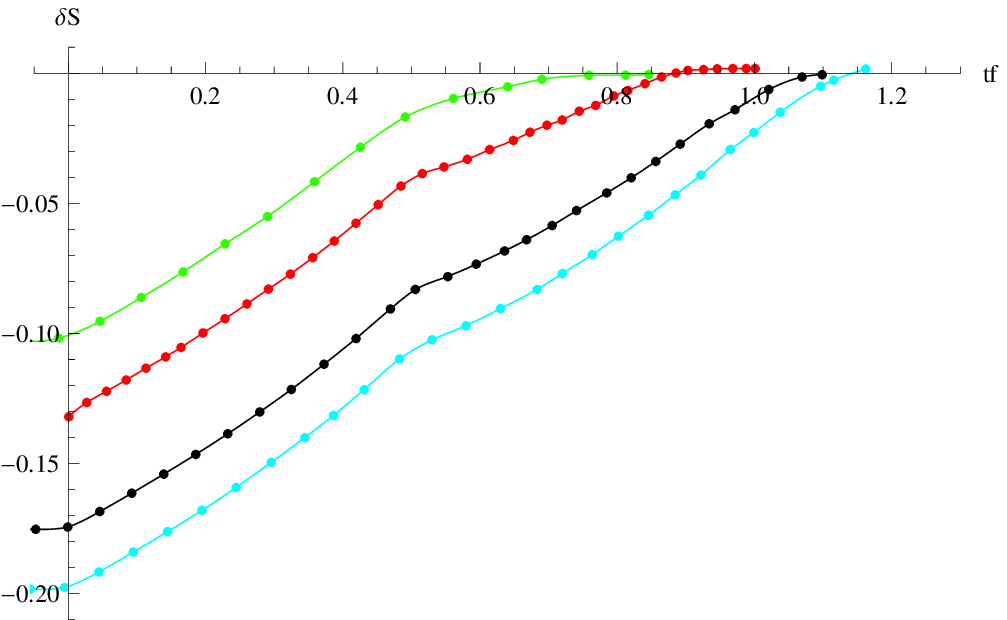}
\caption{
Two point functions of charged operators as a function of $t_f$ for a thermal and chemical potential
quench modeled by an interpolation between $M_{\sf in}=1+Q_{\sf in}^2$ in the far past to $M=3$ in the future,
keeping the charge $Q=Q_{\sf in}=1$ constant (Sect IV.C).
The plots correspond to $d=3$, $\Delta t=0.5$ and $\ell=1,1.4,1.8,2$ (green, red, black and azure
resp.). \\
{\it Left}: $q_E=1$. {\it Right}: $q_E=-1$.}
\label{chotadejara} \ec
\end{figure}

\begin{figure}[ht]
\bc
\includegraphics[width=7.5cm]{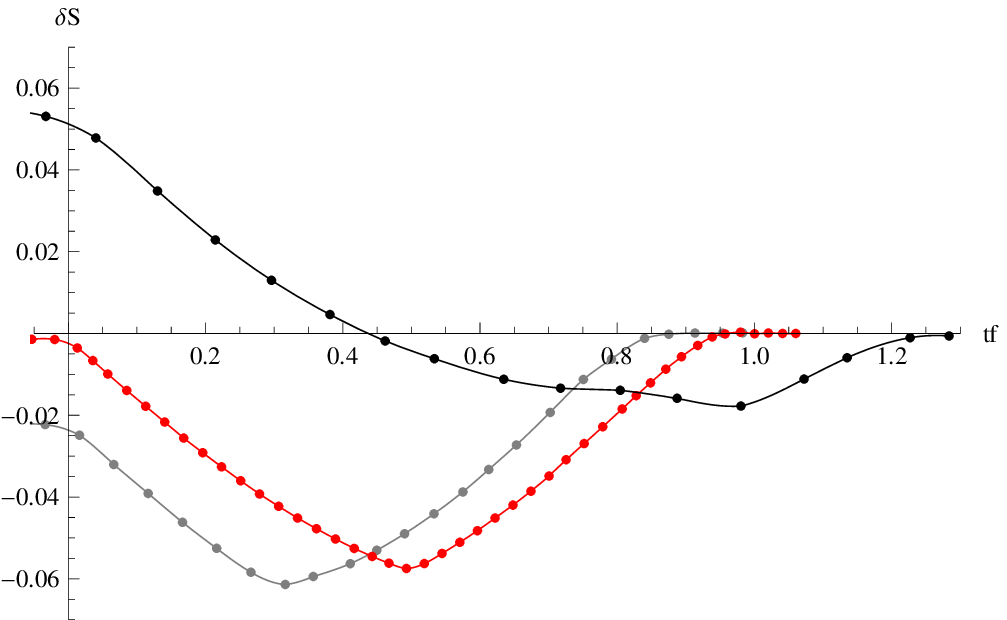}
\includegraphics[width=7.5cm]{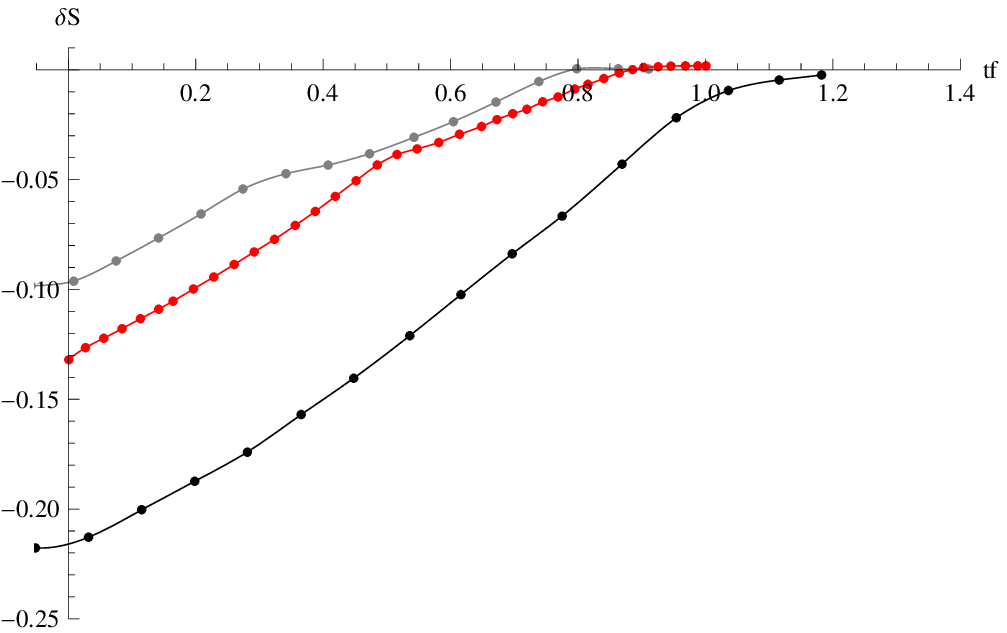}
\caption{
Two point functions of charged operators as a function of $t_f$ for a thermal and chemical potential
quench modeled by an interpolation between $M_{\sf in}=1+Q_{\sf in}^2$ in the far past to $M=3$ in the future,
keeping the charge $Q=Q_{\sf in}=1$ constant (Sect IV.C). The plots correspond to $d=3$, $\ell=1.4$ and
$\Delta t=0.3,0.5,1$ (gray, red and black resp.). {\it Left}: $q_E=1$. {\it Right}: $q_E=-1$.}
\label{chotademontoto} \ec
\end{figure}

\begin{figure}[h]
\bc
\includegraphics[width=7.5cm]{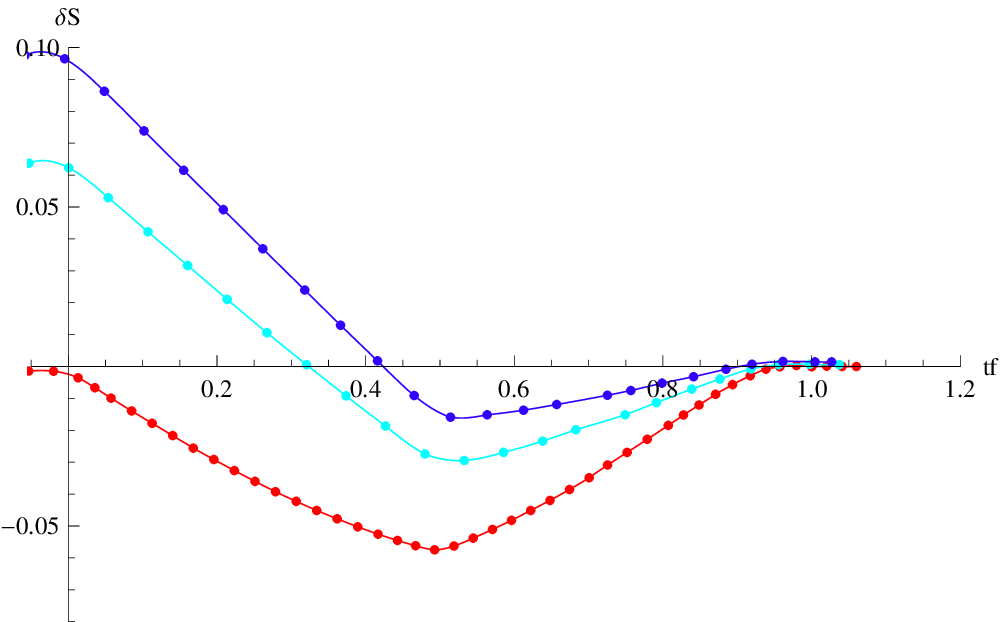}
\includegraphics[width=7.5cm]{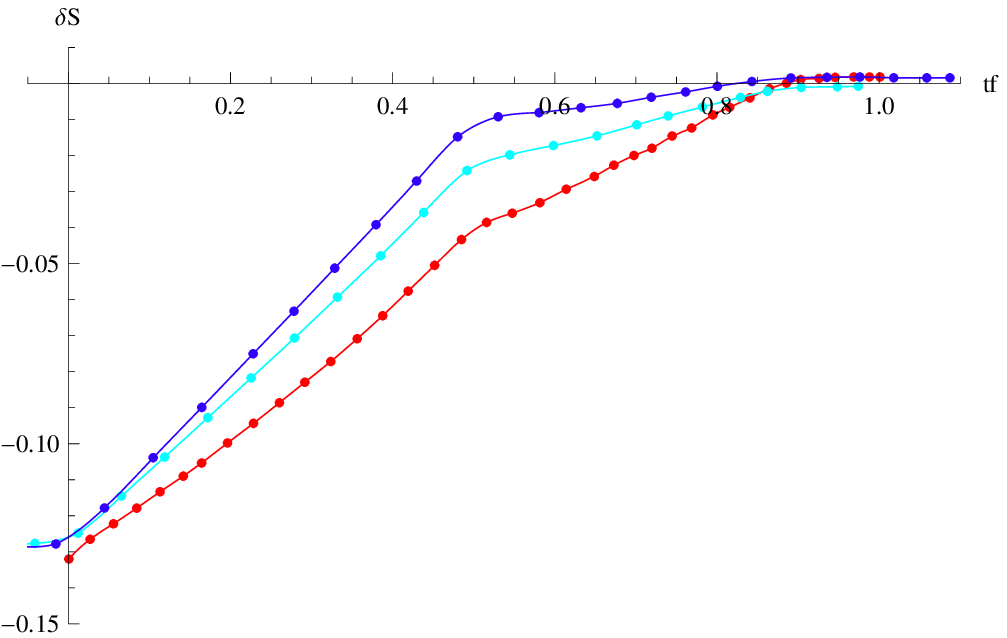}
\caption{
Two point functions of charged operators as a function of $t_f$ for a thermal and chemical potential
quench modeled by an interpolation between $M_{\sf in}=1+Q_{\sf in}^2$ in the far past to $M=3$ in the future,
keeping the charge $Q=Q_{\sf in}=1$ constant (Sect IV.C). The plots correspond to $\ell=1.4$,
$\Delta t=0.5$ and  $d=3,4,5$   (red, azure, blue resp.). {\it Left}: $q_E=1$. {\it Right}: $q_E=-1$.}
\label{chotadelchoto} \ec
\end{figure}

\begin{figure}[h]
\bc
\includegraphics[width=7.5cm]{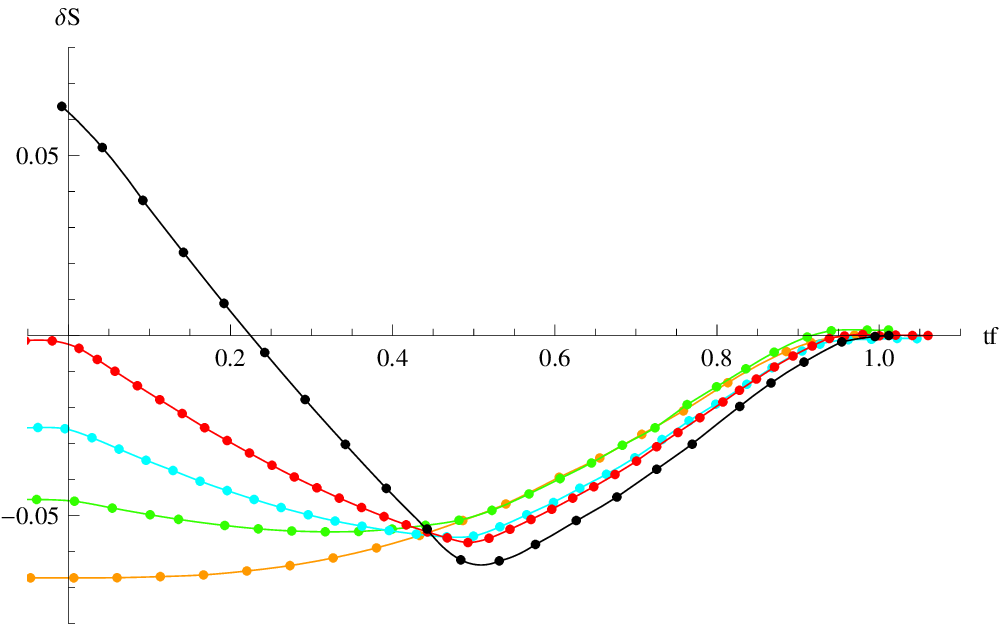}
\includegraphics[width=7.5cm]{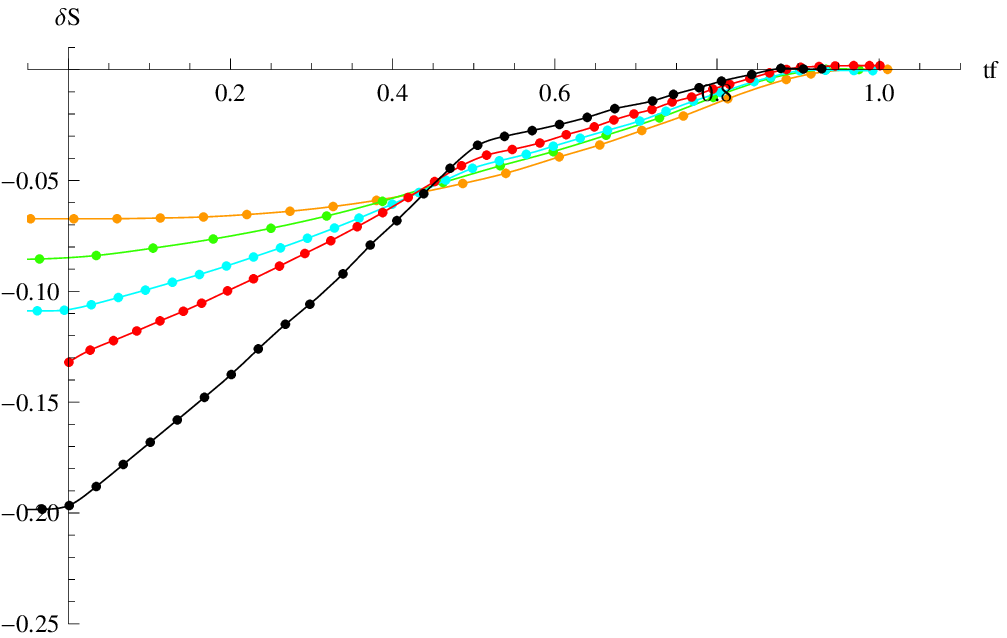}
\caption{
Two point functions of charged operators as a function of $t_f$ for a thermal and chemical potential
quench modeled by an interpolation between $M_{\sf in}=1+Q_{\sf in}^2$ in the far past to $M=3$ in the future,
keeping the charge $Q=Q_{\sf in}=1$ constant (Sect IV.C). The plots correspond to $d=3$, $\ell=1.4$,
$\Delta t=0.5$ and $|q_E|=0,0.3,0.65,1,2$ (yellow,
green, azure, red and black resp.). {\it Left}:  $q_E>0$. {\it Right}: $q_E<0$.}
\label{chotademagoya}
\ec
\end{figure}

Figs. \ref{chotadejara} to \ref{chotademagoya} show the results
corresponding to an interpolation between $M_{\sf in}=2$ and $M=3$ with $Q_{\sf in}=1$.
In Figs. \ref{chotadejara} and \ref{chotademontoto} we see that thermalization time increases as a
function of $\ell$ and $\Delta t$ respectively, showing again the
top-down thermalization effect, and Fig. \ref{chotadelchoto} shows
that the thermalization time increases with the spacetime
dimension. Finally in Fig. \ref{chotademagoya} we see that, for positive chemical potential, the 
thermalization time probed with charged operators grows with the charge $q_E$ of the operator, this is 
an expected result from the gauge theory perspective.

A peculiar feature appears in all the figures: a peak arises at fixed $t_f$
where the derivative of $\delta S$ with respect to $t_f$ has a
sudden change. This can be understood with the help of Fig.
\ref{chotademontoto}, in which it is evident that such peak happens for
$t_f\simeq\Delta t$. Indeed, each point $t_f$ in the curves
represents a geodesic that, according to our boundary conditions
(\ref{boundcond2}), starts at the boundary at $t=t_f-\Delta t$ and
ends at $t=t_f>0$. Since the shell enters space at $t=0$,  {\it early} trajectories
starting at negative times ($t_f<\Delta t$) cross
the shell once in order to return to the boundary. On the other
hand, {\it late} trajectories starting at positive times ($t_f>\Delta t$)
either cross the shell twice, or do not cross it at all, this last
case corresponds to a thermalized situation (see Fig.\ref{lapizjapones2}).
These three classes
of trajectories prove the spacetime in different ways.
 Let us first assume $\hat Qq_E>0$, then, for {\it early} trajectories,
the gravitational force of the background competes with the
electromagnetic interaction during the first part of the trajectory (close to $t=t_f-\Delta t$),
after the particle crosses the shell the two forces pull in the same direction
(close to $t=t_f$). For {\it late} trajectories, the two forces compete only close to
the tip of the trajectory, and cooperate at both extremes, close to $t=t_f-\Delta
t$ and $t=t_f$. Finally in the third case the forces cooperate all
along the trajectory. The same reasoning can be repeated for $\hat
Qq_E<0$, with the regions in which the forces compete or cooperate being
interchanged. This can be visualized in Figs.
\ref{lapizjapones2} and \ref{lapizjapones}, in which the three types of
geodesics are shown. The absence of peaks for vanishing probe charge is
another evidence of their relation to the electromagnetic interaction.

As can be seen in Fig. \ref{chotademagoya}, there is a remarkable
additional feature on the plots:  there exists a value of $t_f$ such that the
function $\delta S$ does not depend on the probe charge $q_E$. We do
not have an explanation for this behavior at the moment.
%
%
%
\subsection{Discharging background}
\label{discharging}
%
To study a discharging background of the kind described in section
\ref{timedep-bulk}  we need to choose $|Q|<|Q_{\sf in}|$. For
illustrative purposes we will take the extreme case $Q=0$ with the
functions $\hat M$ and $\hat Q$ reading
\ba \hat M&=&\frac{M-M_{\sf
in}}{2}\left(1+\tanh\left(\frac{v}{v_0}\right)\right)+M_{\sf in}\,,
\\
\hat Q&=&-\frac{Q_{\sf
in}}{2}\left(1+\tanh\left(\frac{v}{v_0}\right)\right)+Q_{\sf in}\,.
\ea
These functions interpolate between an AdSRN black hole with mass
$M_{\sf in}$ and charge $Q_{\sf in}$ in the distant past $v\ll v_0$
and an AdS black hole with mass $M$ (and vanishing charge) in the
distant future $v \gg v_0$. We again re-scale the radial coordinate
so that $M_{\sf in}=1+Q_{\sf in}^2$  in the initial state. In the
final state, the bound \eqref{hori} is satisfied since the charge vanishes. From
the dual point of view, we are modeling the process of a sudden
decrease of the absolute value of the chemical potential while
energy is being injected into the system. We would like to mention that this
instance cannot be modeled with a constant charge background and complements the
results of the previous section.

\begin{figure}[h]
\bc
\includegraphics[width=7.5cm]{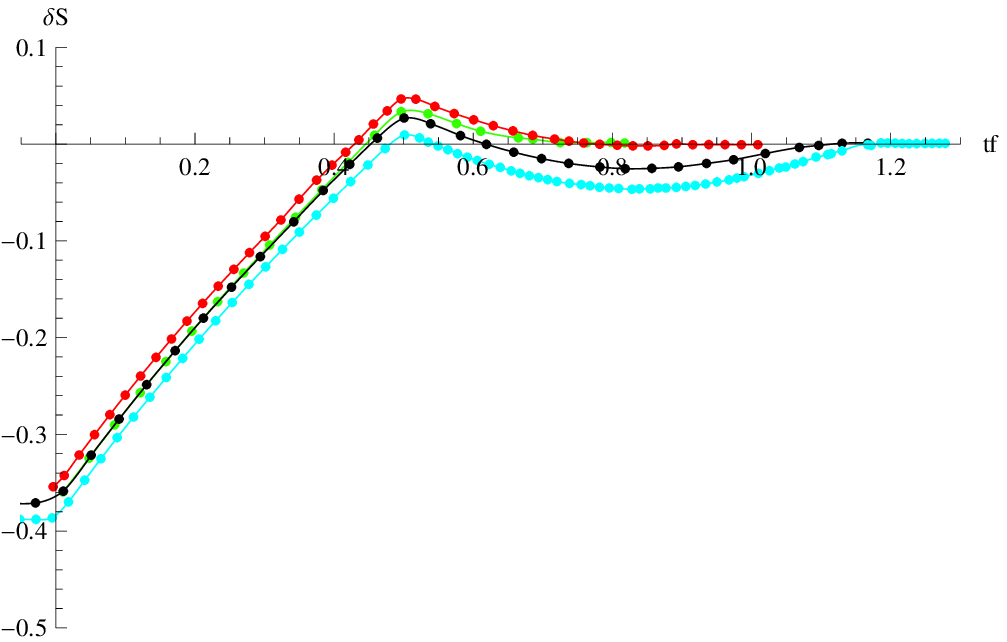}
\includegraphics[width=7.5cm]{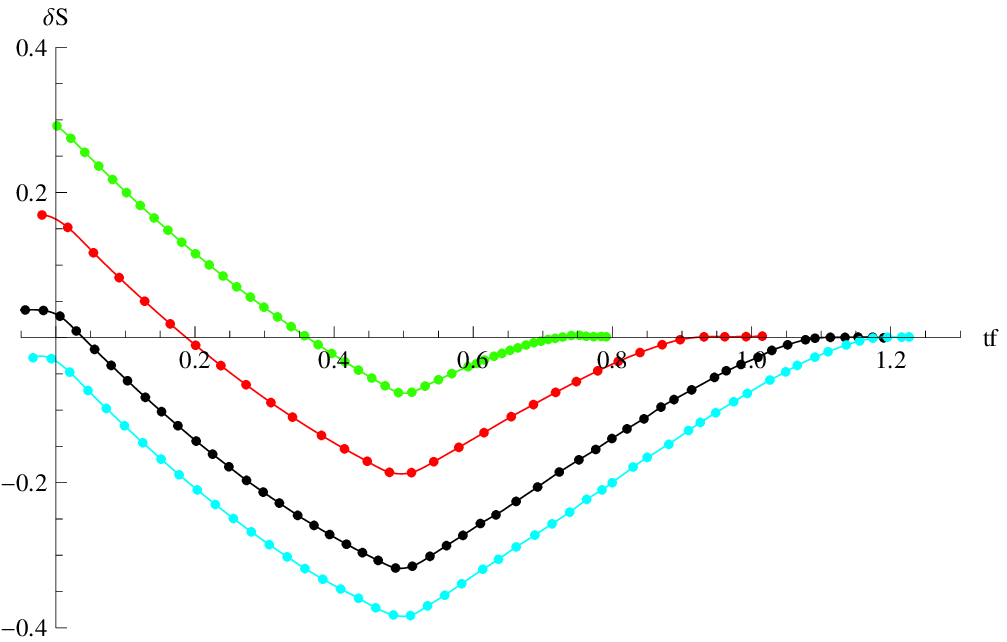}
\caption{ Two point functions of charged operators as a function of $t_f$ for a thermal and chemical potential
quench leading to a final vanishing chemical potential (Sect IV.D). Plots correspond to
$d=3$, $\Delta t=0.5$ and  $\ell=1,1.4,1.8,2$ ( green, red, black and azure
resp.).   {\it Left}: $q_E=1$. {\it Right}: $q_E=-1$. 
In Figs. 10-14, the geometry interpolates between a  RNAdS with $M_{\sf in}=1+Q_{\sf in}^2$, $Q_{\sf in}=-1$ and pure AdS with $M=3$.} \label{chotex}
\ec
\end{figure}

\begin{figure}[h]
\bc
\includegraphics[width=7.5cm]{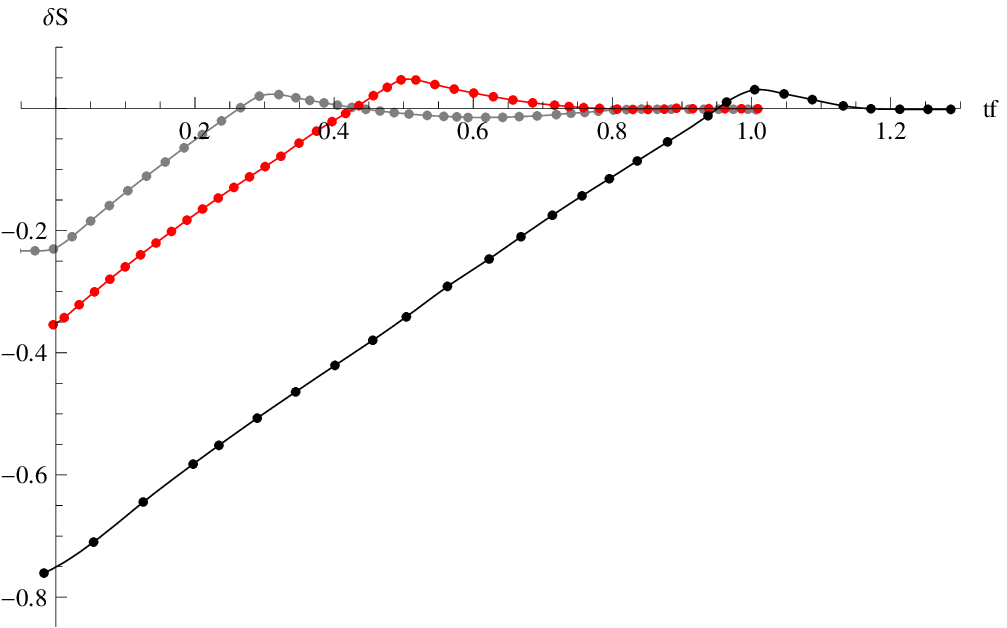}
\includegraphics[width=7.5cm]{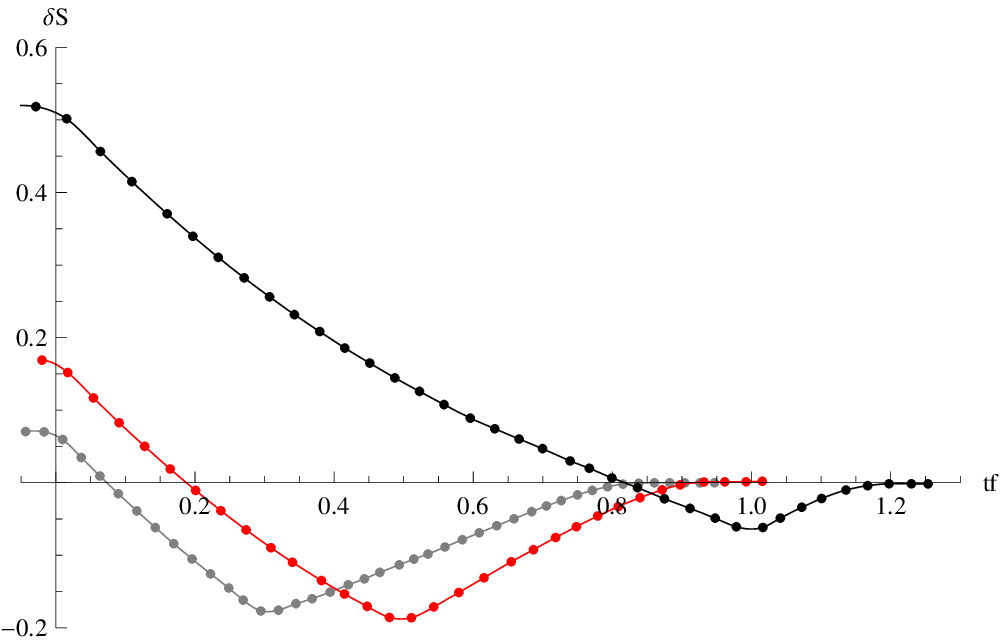}
\caption{Two point functions of charged operators as a function of $t_f$ for a thermal and chemical potential
quench leading to a final vanishing chemical potential (Sect IV.D).
Plots correspond to $d=3$, $\ell=1.4$ and $\Delta t=0.3,0.5,1$ (gray, red and black
resp.). {\it Left}: $q_E=1$. {\it Right}: $q_E=-1$. } \label{chotax}
\ec
\end{figure}

\begin{figure}[h]
\bc
\includegraphics[width=7.5cm]{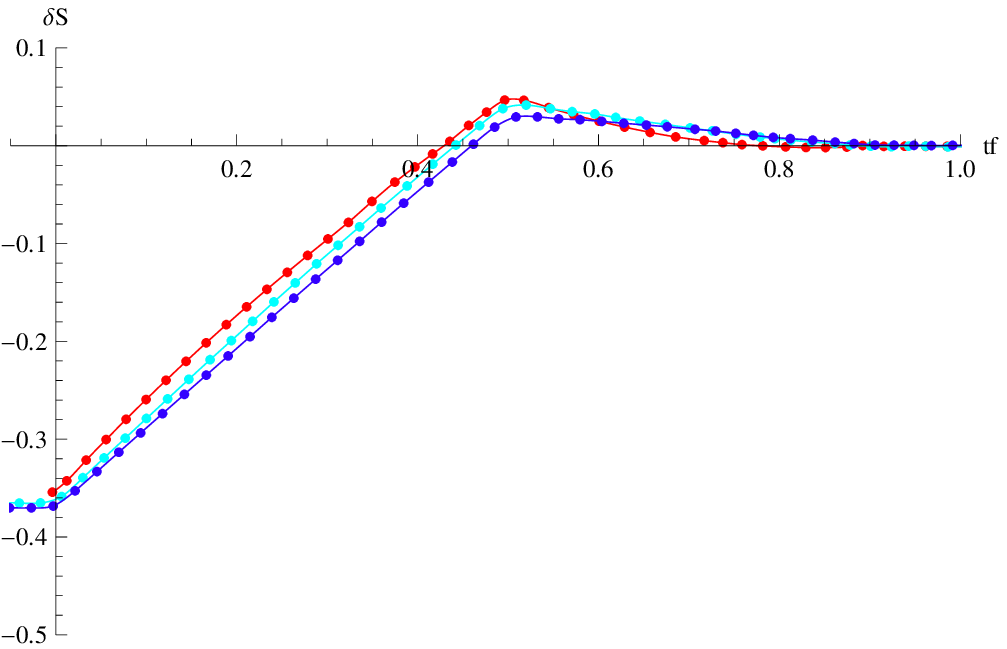}
\includegraphics[width=7.5cm]{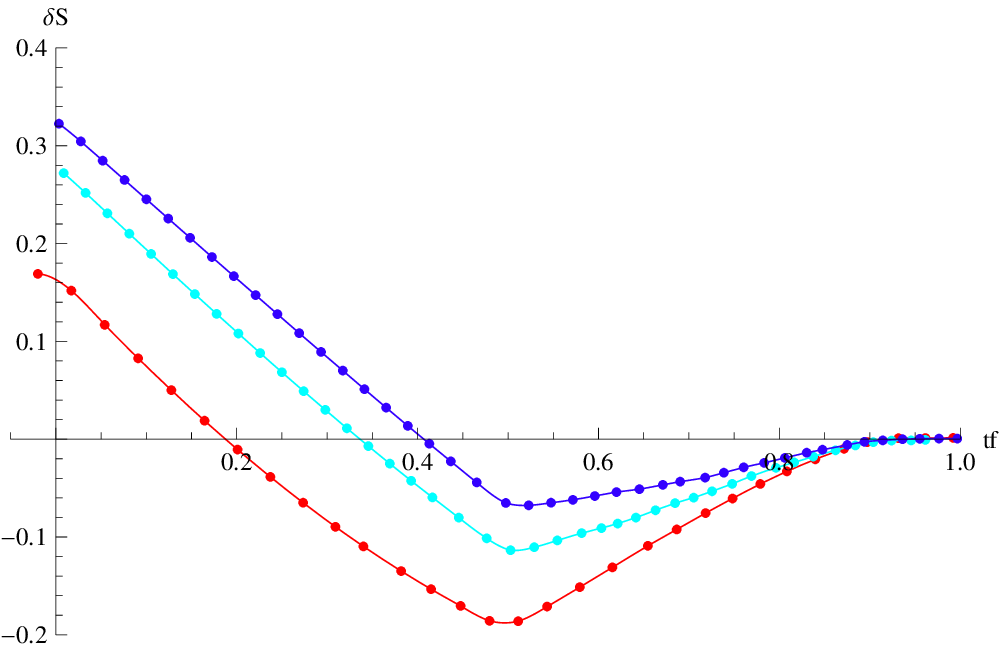}
\caption{Two point functions of charged operators as a function of $t_f$ for a thermal and chemical potential
quench leading to a final vanishing chemical potential (Sect IV.D).
Plots correspond to  $\ell=1.4$, $\Delta t=0.5$ and $d=3,4,5$ (red, azure and blue resp.).
  {\it Left}: $q_E=1$. {\it Right}: $q_E=-1$. }
\label{choton} \ec
\end{figure}

\begin{figure}[h]
\bc
\includegraphics[width=7.5cm]{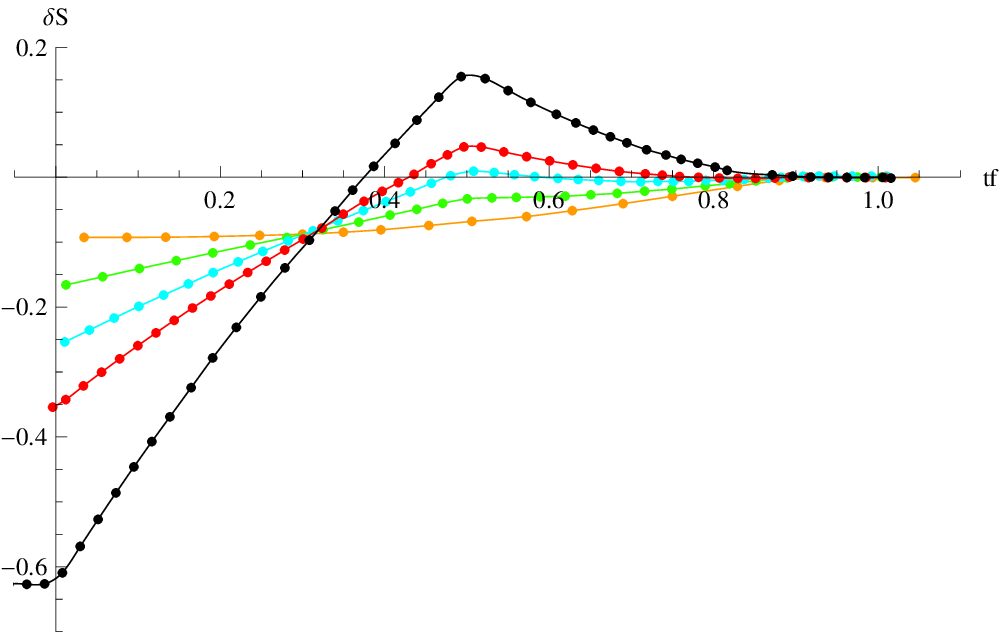}
\includegraphics[width=7.5cm]{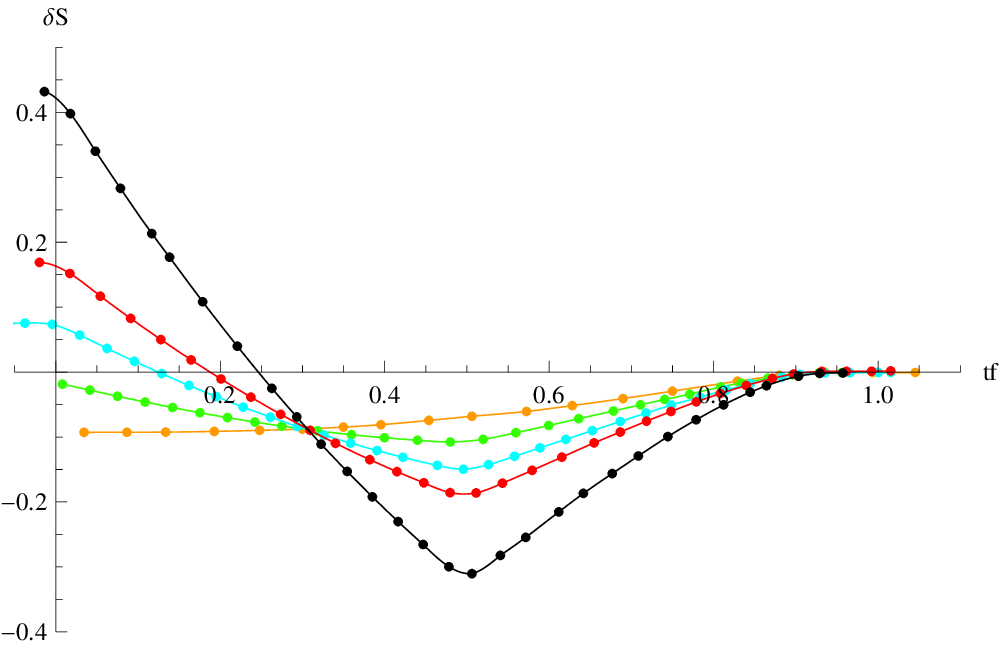}
\caption{ Two point functions of charged operators as a function of $t_f$ for a thermal and chemical potential
quench leading to a final vanishing chemical potential (Sect IV.D).
Plots correspond to $d=3$, $\ell=1.4$, $\Delta t=0.5$ and $|q_E|=0,0.3,0.65,1,2$ (orange,
green, azure, red and black resp.).  
{\it Left}:  $q_E>0$. {\it Right}: $q_E<0$.  } \label{chotin} \ec
\end{figure}

Results are shown in Figs. \ref{chotex} to \ref{chotin}. Again we
see that top-down thermalization arises and that the thermalization time
grows with $\ell$, $\Delta t$ and the space dimension.
The peak at $t_f=\Delta t$ is also present, the reasons being the same
as explained in the previous section. Thermalization time also increases with the
charge of the probe.
%

%
%
%
%
%
%
\subsection{Charging background}
\label{charging}
%
%
%
%
We conclude by analyzing the thermalization process for a quench leading to an increase
on both the temperature and the chemical potential (this situation has been considered before in
\cite{martin} for the case of uncharged operators). The functions $\hat M$ and $\hat Q$ read
\ba \hat
M&=&\frac{M}{2}\left(1+\tanh\left(\frac{v}{v_0}\right)\right)\,,
\\
\hat
Q&=&\frac{Q}{2}\left(1+\tanh\left(\frac{v}{v_0}\right)\right)\,. \ea
These functions interpolate between a pure AdS solution in the
distant past $v\ll v_0$, to a AdSRN solution in the distant future
$v\gg v_0$ with mass $M$ and charge $Q$. We rescale
the radial coordinates so as to have $M=1+Q^2$ in the future state.
From the dual point of view, as energy flows into the system, the absolute
value of the chemical potential suddenly increases.
As discussed in Sect \ref{timedep}, in the present situation the null
energy condition is violated in the deep IR, so we must check that
our geodesics do not reach such region (see Fig. \ref{japi}).

\begin{figure}[h]
\bc
\includegraphics[width=7.5cm]{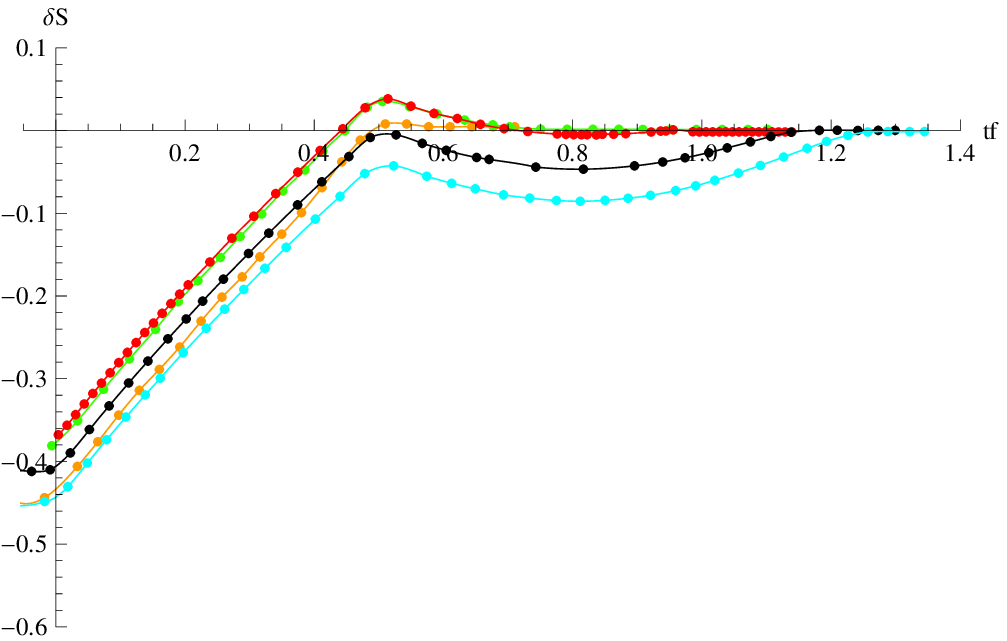}
\includegraphics[width=7.5cm]{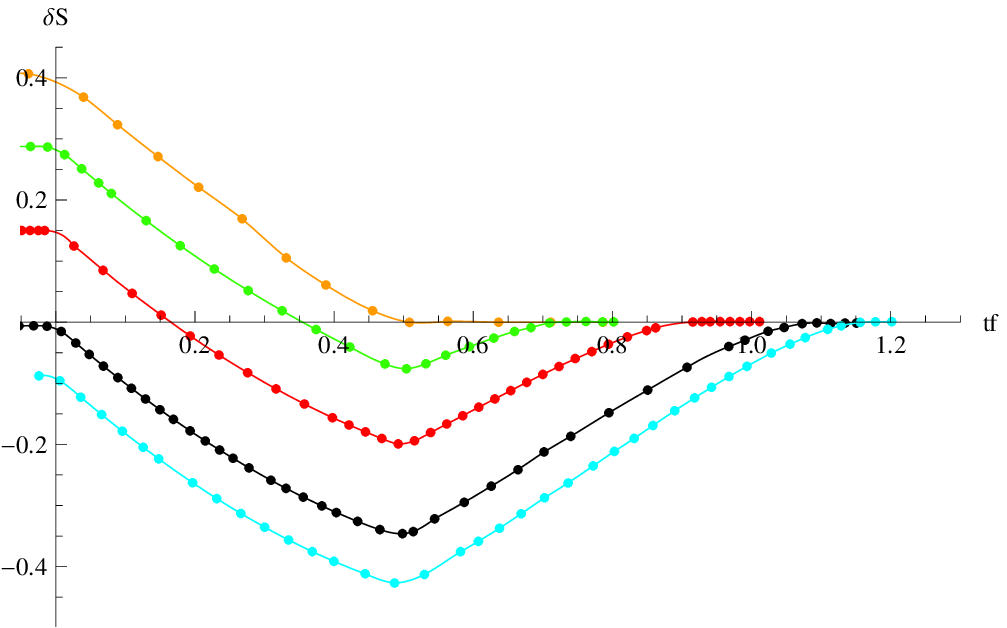}
\caption{ Two point functions of charged operators as a function of $t_f$ for a thermal and chemical potential
quench modeled  by a charging background (Sect. IV.E). Plots corresponding to
$d=3$, $\Delta t=0.5$ and $\ell=1,1.4,1.8,2$ (orange, green, red, black and azure
resp.). {\it Left}: $q_E=1$. {\it Right}: $q_E=-1$. \\
The geometry interpolates between pure AdS and AdSRN with  $M=1+Q^2$ and $Q=1$. } \label{chotor}
\ec
\end{figure}

\begin{figure}[ht]
\bc
\includegraphics[width=7.5cm]{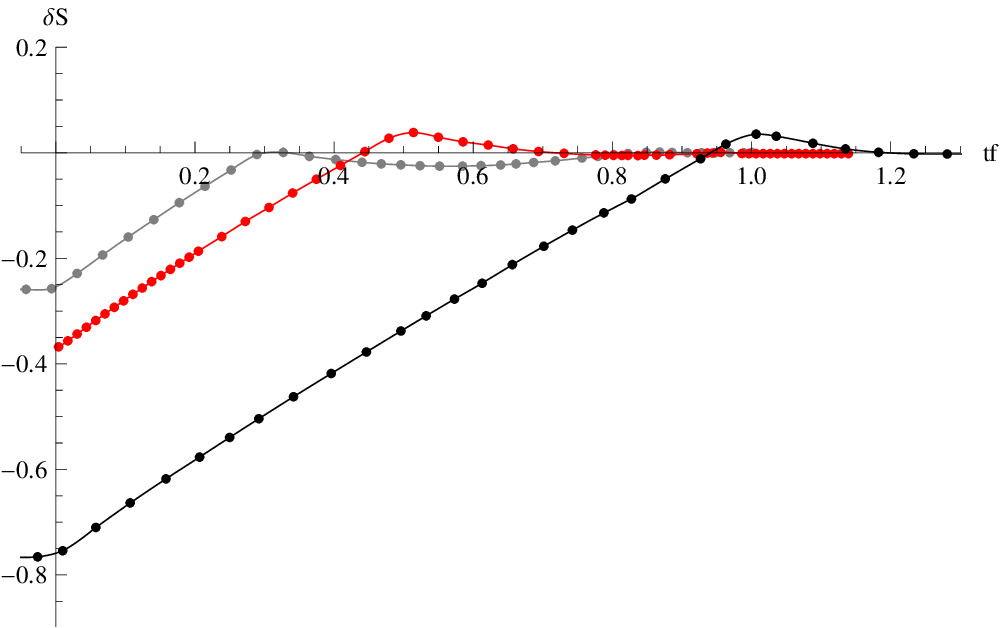}
\includegraphics[width=7.5cm]{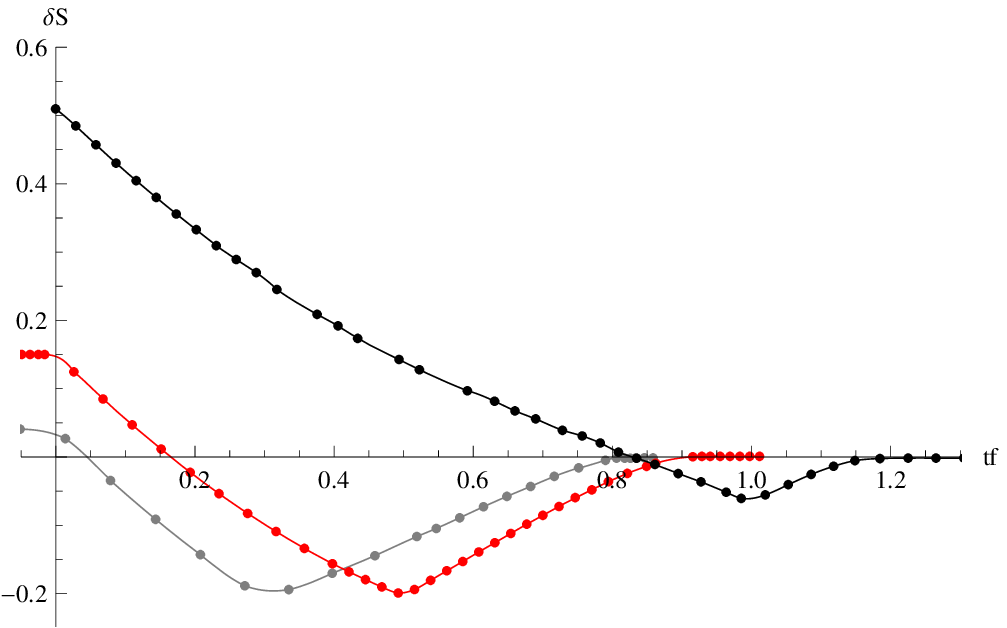}
\caption{ Two point functions of charged operators as a function of $t_f$ for a thermal and chemical potential
quench modeled  by a charging background (Sect. IV.E). Plots corresponding to
$d=3$, $\ell=1.4$ and $\Delta t=0.3,0.5,1$ (gray, red and black resp.).\\
{\it Left}: $q_E=1$. {\it Right}: $q_E=-1$.  }
\label{choter} \ec
\end{figure}

\begin{figure}[ht]
\bc
\includegraphics[width=7.5cm]{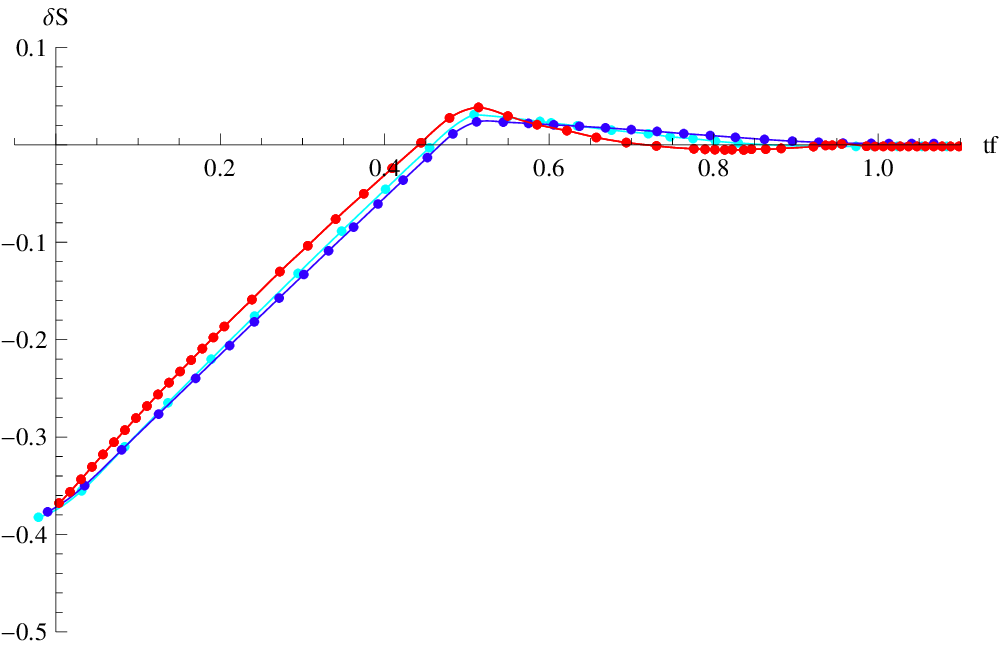}
\includegraphics[width=7.5cm]{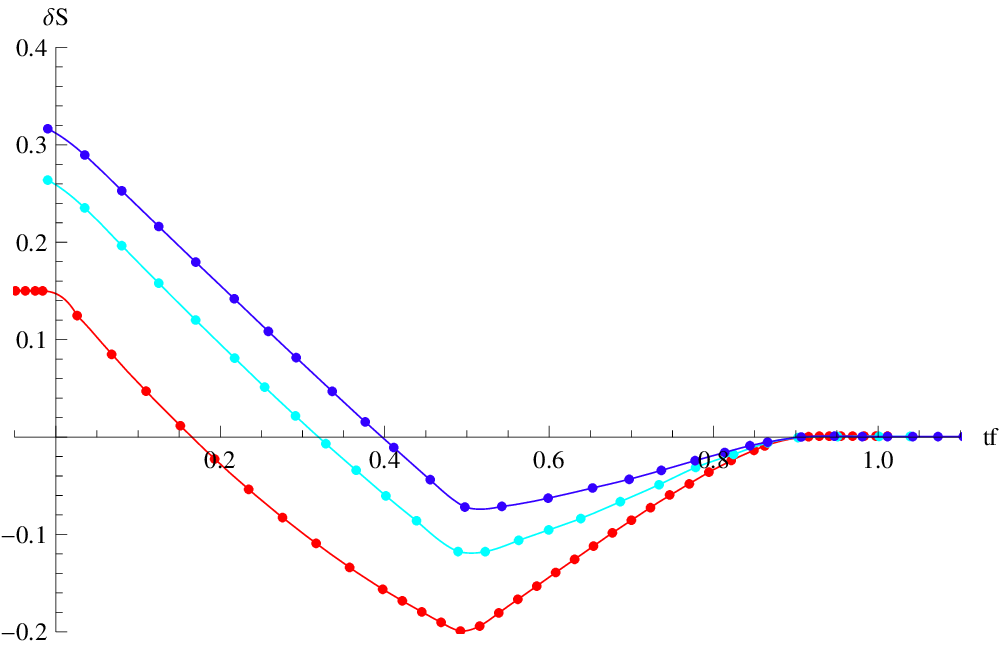}
\caption{Two point functions of charged operators as a function of $t_f$ for a thermal and chemical potential
quench modeled  by a charging background (Sect. IV.E). Plots corresponding to
$d=3$, $\ell=1.4$ and $\Delta t=0.5$ for $d=3,4,5$ (red, azure and  blue resp.).
{\it Left}: $q_E=1$. {\it Right}: $q_E=-1$. }
\label{chotir} \ec
\end{figure}

\begin{figure}[ht]
\bc
\includegraphics[width=7.5cm]{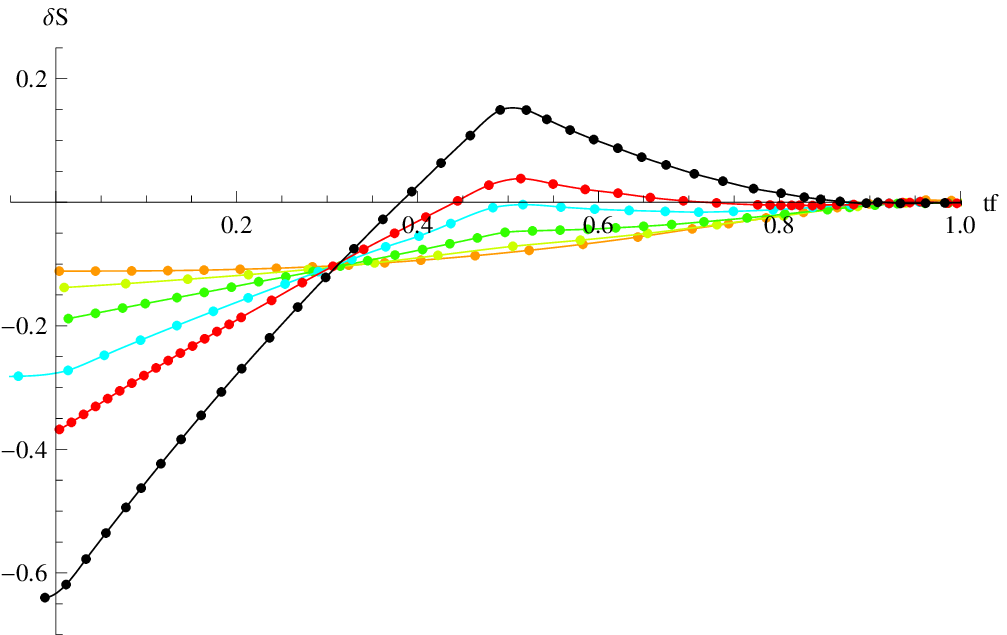}
\includegraphics[width=7.5cm]{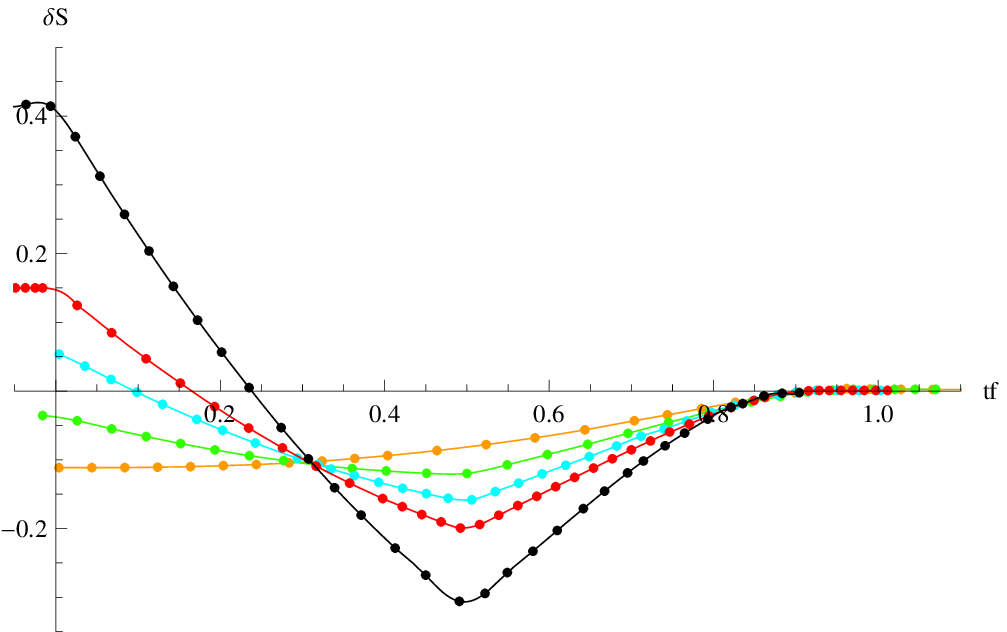}
\caption{Two point functions of charged operators as a function of $t_f$ for a thermal and chemical potential
quench modeled  by a charging background (Sect. IV.E). Plots corresponding to
$d=3$, $\ell=1.4$ and $\Delta t=0.5$.  {\it Left}: $q_E=0,0.1,0.3,0.65,1,2$ (orange, light yellow, green,
 azure, red, black resp.)
{\it Right}: $-q_E=0,0.3,0.65,1,2$ (orange, green, azure, red and black resp.). } \label{chotur}
\ec
\end{figure}

\begin{figure}[ht]
\bc
\includegraphics[width=7.5cm]{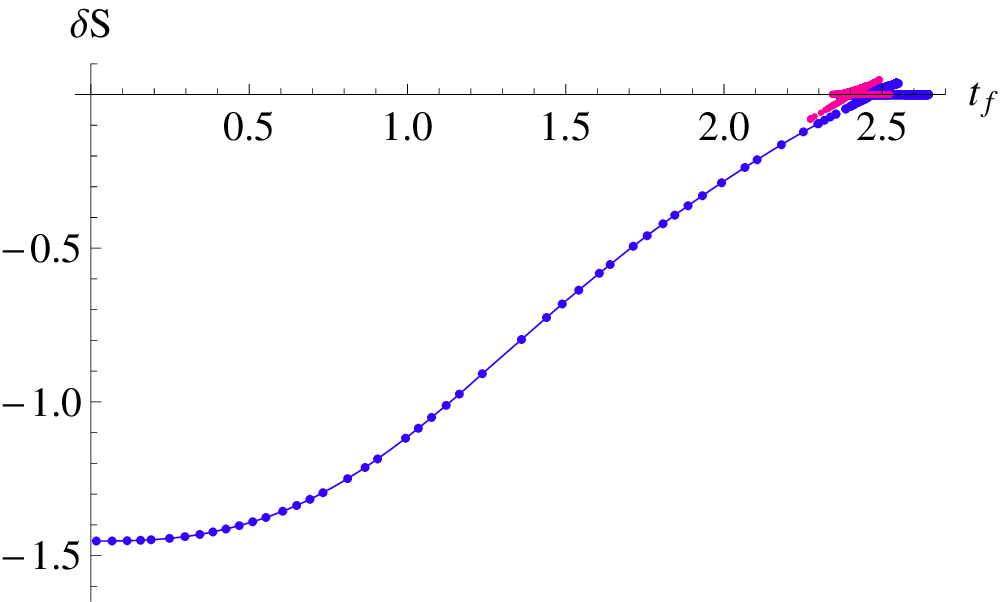}
\includegraphics[width=7.5cm]{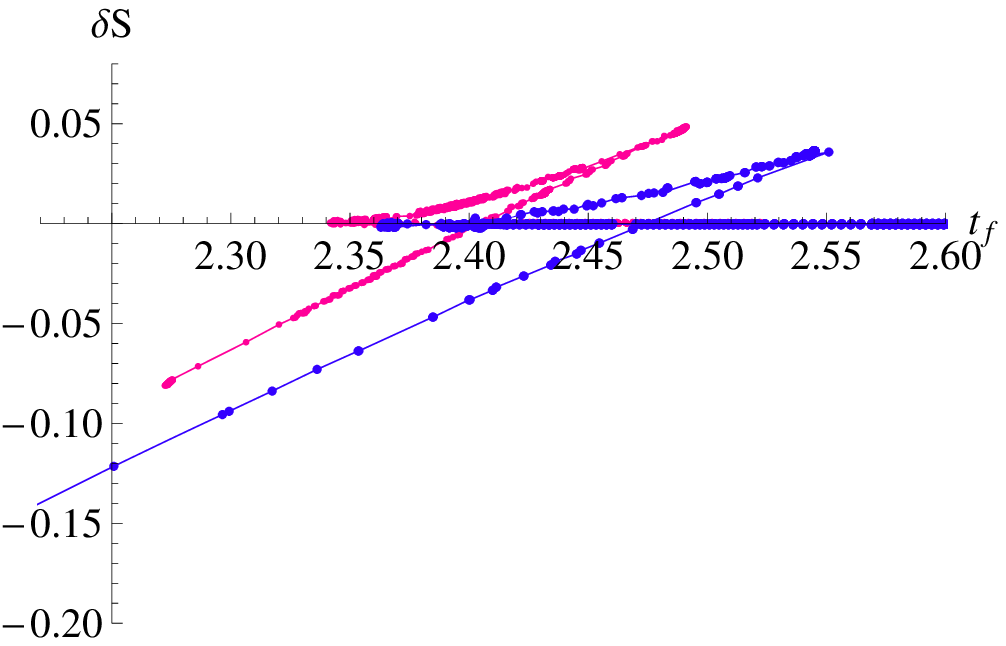}
\caption{Two point functions of charged operators as a function of $t_f$ for a thermal and chemical potential
quench modeled  by a charging background (Sect. IV.E). Plots corresponding to
a charging backgound from pure AdS to an extremal AdSRN. Plots coreespond to $d=3$, $M=1+Q^2$, $Q=\sqrt3$,
$\ell=4$, $\Delta t=0.5$ and $q_E=0,0.5$ (blue and violet resp.). On the right we have zoomed the swallow tale
region.} \label{choturex} \ec
\end{figure}

Results are shown in Figs. \ref{chotor} to \ref{chotur}, with
features similar to the previous cases, namely, top-down
thermalization, thermalization time growing with the dimension of
space, peaks denoting the transition between the different classes of
geodesics, etc. It is worth mentioning that the swallow tale
structure found in \cite{martin} also appears for non-vanishing
probe charges as can be seen in Fig. \ref{choturex}.


%

~

~

~

\section{Conclusions}
\label{conclusion}
%

We have analyzed the  energy conditions for the external matter
needed to support a family of charged AdS-Vaidya metrics. The
metrics studied interpolate between two AdSRN black holes with
different mass and charge, or between pure AdS and an AdSRN black
hole with non-vanishing charge. They have been used in the
literature,  via the AdS/CFT correspondence, to model the
thermalization process in a strongly coupled plasma after a quench
in energy and chemical potential.

We found that the null energy condition is violated in the infrared
region of the geometry for increasing mass whenever the absolute
value of the black hole charge increases in time. On the other hand,
when the absolute value of the black hole charge is kept constant or
decreases, the null energy condition is satisfied everywhere.
This implies that charged Vaidya metrics can be used to analyze thermalization
processes for all energy scales only when the quench decreases the absolute
value of the chemical potential. On the other
hand, when the quench increases the absolute value of the chemical
potential, then the metric is only useful for probing the thermalization process
above an IR cutoff1.

We applied the above results to study the thermalization of a
strongly coupled plasma after a quench in the energy and chemical potential, considering the cases
where the chemical potential either increases or decreases in absolute value. As probe of
thermalization we considered charged operators two point functions. We found
that the thermalization time increases with the charge of the operator, as well as with the dimension of
the field theory. As expected in these kind of holographic
constructions, the thermalization is top-down, in the sense that UV
degrees of freedom thermalize earlier, followed by IR ones.

%
%
%
%
%
%
\section{Acknowledgments}
%
%
%
%
We thank Martin Schvellinger, Walter Baron and Damian Galante for
helpful discussions on thermalization of strongly coupled plasmas
via AdS/CFT. We also thank Anibal Iucci and Nicol\'as Nessi for
helpful exchange on the general knowledge about thermalization after
quenches and Jer\'onimo Peralta Ramos and Pablo Rodriguez Ponte for
discussions. This work was partially supported by Conicet grant
PIP2009-0396, PIP 0595/13, ANPCyT grant PICT2008-1426 and UNLP
grant 11/X648.
\newpage
\appendix
%
%
%
%
%
%
\section{Eddington-Finkelstein null coordinates}
\label{EF}
%
%
%
For completeness we quote here some well known facts of the
Eddington-Finkelstein coordinate system $X^\mu=(v,{\bf x},z)$ chosen
in \eqref{RNADSBH}. Parametrizing the geodesics as $X^\mu=(v(z),{\bf
x}(z),z)$ it is immediate to see that the curve $X^\mu=(v_0,{\bf
x}_0,z)$ moving along the holographic direction is null. In what
follows we show that the sign of the $dvdz$ term in \eqref{RNADSBH}
determines whether the curve is either ingoing or outgoing. This in
turn determines whether the mass shell in a Vaydia metric is ingoing
or outgoing.

We start by considering the timelike vector $\partial_v$ to be
future directed. The null geodesics on the $(v,z)$ plane for the
AdS$_{d+1}$ black hole geometry are obtained from
\be (1-Mz^d)dv^2\pm2dvdz=0\Longrightarrow
v(z)=C\mp2z\,_2F_1\left(1,\frac1d,1+\frac1d,Mz^d\right)\,,
\label{EFgeod} \ee
here $C$ is an integration constant and $_2F_1$ is Gauss
hyper-geometric function, blowing up at $Mz^d=1$ and having an
expansion $_2F_1\approx  z+{\cal O}(z^{d+1})$ near the boundary of
AdS. The upper sign choice in \eqref{EFgeod} therefore implies that
the geodesic displayed in \eqref{EFgeod} is escaping from the
horizon as $v$ increases. Taking into account that $\partial_v$ is
timelike outside the horizon, we conclude that the $v=v_0$ curve
corresponds to a radially ingoing null geodesic.

\begin{figure}[h]
\bc
\includegraphics[width=6.5cm]{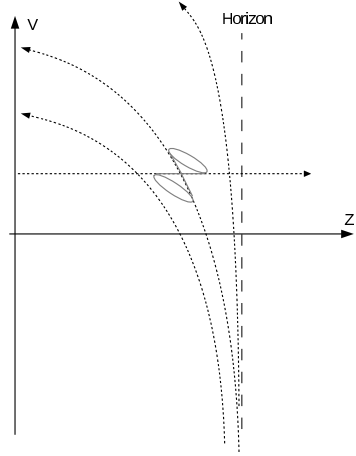}
\vspace{-1cm} \caption{Null geodesics and light cones for the EF
(ingoing) coordinates chosen in the body of the paper. The plot
shows the solutions of \eqref{EFgeod} for the upper sign.}
\label{horizon} \ec
\end{figure}
%
%
%
\section{World line formalism and geodesic approximation}
\label{Worldline}
%
%

%
The bulk propagator $G(x_2^\alpha,z_2 |\,x_1^\alpha,z_1)$ in
(\ref{correlator}) is the Green function of the equation of motion
of a bulk charged scalar field, or in other words
\ba G(x_2^\alpha,z_2 |\,x_1^\alpha,z_1)&=&\left\langle
{x_2^\alpha,z_2}\left| \,\frac {i} {g^{\mu\nu}D_\mu
D_\nu-m^2+i\epsilon}\, \right|{x_1^\alpha,z_1}\right\rangle\,,
\label{propa} \ea
where $D_\mu=\partial_\mu-i {\sf e} A_\mu$ is the covariant
derivative for a charged scalar of charge ${\sf e}$.    Schwinger's
proper time representation consists in rewriting the inverse
operator \eqref{propa} as \cite{bastianelli,Marolf,kiri}
\ba G(x_2^\alpha,z_2|\,x_1^\alpha,z_1)&=& \int_0^\infty d T
\left\langle {x_2^\alpha,z_2}\left| \,e^{iT \left(g^{\mu\nu} D_\mu
D_\nu-m^2+i\epsilon\right)}\,
\right|{x_1^\alpha,z_1}\right\rangle\,. \ea
The exponential inside the bracket can be understood as the
evolution operator for the Hamiltonian $H=-g^{\mu\nu}D_\mu
D_\nu+m^2$, which allows us to write
\ba
&&G({x_2^\alpha,z_2|\,x_1^\alpha,z_1})=\int_0^\infty dT \int_{x
(0)=(x_1^\alpha,z_1)}^{x^\mu(T)=(x_1^\alpha,z_1)} \!\!\!\!{\cal
D}x(\tau) \, e^{iS_{\sf 1particle}[x(\tau)] } \,,
\n
&&\mbox{with}\qquad
S_{\sf 1particle}[x(\tau)]= \int_0^T d\tau \left( \frac14 g_{\mu\nu}
\dot x^\mu \dot x^\nu + {\sf e} A_\mu \dot {x}^\mu -m^2\right)\,.
\ea
Here $S_{\sf 1particle}[x(\tau)]$ is a one-particle action written
in terms of a world line proper time parameter $\tau\in[0\dots T]$,
and a dot ($\dot~$) means derivative with respect to $\tau$. We can
rescale $\tau \to \tau T$ in order to get $\tau \in[0\dots1]$ and we
end up with
\ba &&G({x_2^\alpha,z_2|\,x_1^\alpha,z_1})=\int_0^\infty dT
\int_{x^\mu(0)=(x_1^\alpha,z_1)}^{x^\mu(1)=(x_2^\alpha,z_2)}
\!\!\!\!{\cal D}x(\tau) \, e^{iS_{\sf 1particle}[x(\tau)] } \,,
\n
&&\mbox{with}\qquad S_{\sf 1particle}[x(\tau)]= \int_0^1 d\tau
\left( \frac1{4T} g_{\mu\nu} \dot x^\mu \dot x^\nu + {\sf e} A_\mu
\dot{x}^\mu -T m^2\right)\,. \label{original} \ea
Notice that this one-particle action is not invariant under
reparametrization of the world line. Introducing the einbein
$e_\tau(\tau)$, we can interpret (\ref{original}) as a gauge fixed
expression for an originally reparametrization invariant action
\cite{Marolf,kiri}\footnote{ Action \eqref{repa-invariant} is
invariant under local reparametrizations $\delta\tau=\xi(\tau)$ by
virtue of $\delta e_\tau=-\partial_\tau(e_\tau\xi)$.}
\ba &&G({x_2^\alpha,z_2|\,x_1^\alpha,z_1})=\int {\cal
D}e_\tau(\tau)\delta({\dot e_\tau}) \int_{x^\mu_{\sf
initial}=(x_1^\alpha,z_1)}^{x^\mu_{\sf final}=(x_2^\alpha,z_2)}
\!\!\!\!{\cal D}x(\tau) \, e^{iS_{\sf invariant}[x(\tau)] } \,,
\n
&&\mbox{with}\qquad S_{\sf invariant}[x(\tau)]= \int d\tau \left(
\frac1{4e_\tau} g_{\mu\nu} \dot x^\mu \dot x^\nu + {\sf e} A_\mu
\dot{x}^\mu -e_\tau m^2\right)\,. \label{repa-invariant} \ea
The  gauge fixing amounts to setting $e_\tau =T$, and the $T$
integral in \eqref{original} corresponds to the leftover
(Teichm\"uller) parameter after gauge fixing \cite{kiri}. In the
limit of large mass $m$, in which we are interested, we can take the
semiclassical approximation of the integral (\ref{repa-invariant}),
to have
\ba &&G({x_2^\alpha,z_2|\,x_1^\alpha,z_1})=e^{i S_{\sf on-shell}
(x^\alpha_1,z_1;x^\alpha_2,z_2)} \,,\label{pind} \n &&\mbox{with}\qquad
S_{\sf on-shell} ({x_1^\alpha,z_1|\,x_2^\alpha,z_2})= \left. \int d\tau
\left(-m\sqrt{-g_{\mu\nu}\dot x^\mu \dot x^\nu} + {\sf e} A_\mu\dot
x^\mu\right) \right|_{x^\mu(\tau)=x^\mu_{\sf classical}(\tau)}\,,
\label{action2} \ea
where the quotient ${\sf e}/m$ is assumed finite in the large $m$
limit, and $x^\mu_{\sf classical}(\tau)$ are the classical
trajectories starting at $(x^\alpha_1,z_1)$ and ending at
$(x_2^\alpha,z_2)$.
%
%
%
%
%
%
\section{Accessing the classically forbidden region}
\label{Euclid}
%
%
%
%
The classical trajectories we need to compute lie completely in the
classically forbidden region. Indeed, from the action
\be
S= \int d\tau \left(-m \sqrt{-g_{\mu\nu}\dot x^\mu \dot x^\nu} +  {\sf
e} A_\mu\dot x^\mu\right)\,,
\label{action3}
\ee
we find the canonical momenta
\be p_\mu = \frac {m\,g_{\mu\nu}\dot x^\nu}{\sqrt{-g_{\mu\nu}\dot
x^\mu \dot x^\nu}}+{\sf e}A_\mu\,, 
\label{curvedmomenta}\ee
and time reparametrization invariance imply 
\be g^{\mu\nu}(p_\mu-{\sf e}A_\mu)(p_\nu-{\sf e}A_\nu)+m^2=0\,, 
\label{massconstr}
\ee
As we now show, these momenta become imaginary in the near boundary
region, implying that such region is forbidden from a classical
point of view.
%
%
%
\subsection{Vanishing probe charge}
%
%
Let us first consider the case ${\sf e}=0$. Using the explicit form
of our metric and gauge fields, the action reads
\be S=-mL\int d\tau \,\frac1z\sqrt{-\dot {\bf x}_{d-1}^2+ f \dot
v^2+2\dot z\dot v}\,. \label{action4} \ee
The resulting canonical momenta are
\ba
p_v&=& -\frac {mL}{zR}(f\dot v+\dot z)\,,
\nonumber\\
p_z&=& -\frac {mL}{zR}\dot v \,,\qquad\qquad\qquad\qquad {\rm
with}\quad R=\sqrt{-\dot {\bf x}_{d-1}^2+ f \dot v^2+2\dot z\dot v}\,,
\nonumber\\
{\bf p}_{d-1} &=& \frac {mL}{zR} \dot {\bf x}_{d-1}\,, \ea
and the constraint \eqref{massconstr} results
\be f p_z^2-2p_v p_z+{\bf p}^2_{d-1}=-\frac{m^2L^2}{z^2}\,.
\label{rel2} \ee
Since for small $z$ we have  $f\simeq 1$, any $v$-dependence in $f$ is washed 
out near the boundary implying that the quantity $p_v$
becomes constant. For fixed $p_v$, the quadratic polynomial in
$p_z$ in the left hand side of \eqref{rel2} has a lower bound at
$p_z=p_v$. On the other hand, the right hand side becomes
arbitrarily negative as $z\to0$. Thus, for $z$ small enough, the
relation cannot be satisfied with real momenta, and the momenta must
become complex. Indeed, for small enough $z$ the relation
\eqref{rel2} can be solved as $p_z \approx imL/z+p_v$,   implying
that it can be satisfied with pure imaginary momenta $p_z=ip_z^E$
and $p_v=i p_v^E$. Based on this, for generic $z$ we propose the
ansatz ${\bf p}_{d-1}=i{\bf p}_{d-1}^E$, $p_z=ip_z^E$ and $p_v=i
p_v^E$, and in this new variables we have
\ba p_v^E&=& i\frac {mL}{zR}(f\dot v+\dot z)\,,\nonumber
\\
p_z^E&=& i\frac {mL}{zR}\dot v \,,\qquad\qquad\qquad\qquad {\rm
with}\quad R=\sqrt{-\dot {\bf x}_{d-1}^2+ f \dot v^2+2\dot z\dot
v}\,,\nonumber
\\
{\bf p}^E_{d-1} &=&- i\frac {mL}{zR} \dot {\bf x}_{d-1} \,,
\ea
and
\be -f{p^E_z}^2+2p^E_v p^E_z-{{\bf
p}^E_{d-1}}^{\!\!\!\!2}=-\frac{m^2L^2}{z^2}\,. \label{rel3}
\ee
We can re-absorb the $i$ factors in the velocities via a Wick
rotation of the wordline time $\tau=-i\tau_E$ obtaining
\ba p_v^E&=& -\frac {mL}{zR}(f v'+z')\,,\nonumber
\\
p_z^E&=& -\frac {mL}{zR} v'\,, \qquad\qquad\qquad\qquad {\rm with}\quad
R=\sqrt{{\bf x}_{d-1}'^2- f v'^2-2 z' v'}\,,\nonumber
\\
{\bf p}^E_{d-1} &=& \frac {mL}{zR} {\bf x}'_{d-1}\,, 
\label{emomenta}
\ea
where a prime $(')$ means derivative with respect to $\tau_E$. With
these redefinitions, the  momenta \eqref{emomenta} can be re-interpreted as derived
from the Euclidean action obtained via the substitution
$\tau=-i\tau_E$ in \eqref{action4}, namely
\be
 S_E=mL\int d\tau_E \,\frac1z\sqrt{{{\bf x}'}_{d-1}^2- f
v'^2-2z'v'}\,. \label{action4E} \ee
In other words, the imaginary momenta of \eqref{action4} in the
forbidden region can be re-interpreted as the real momenta of its
``Euclidean worldline time'' version \eqref{action4E}.

This ``Euclideanization'' implies re-interpreting the complex valued
classical solution of the equations of motion in the classically
forbidden region, as the real valued classical solution of the Wick
rotated system. Notice that in solving relation \eqref{rel2} for
small enough $z$ we could include a real part on the momenta
$p_z=ip_z^E+p_{Re}$ and $p_v=i p_v^E-p_{Re}$, or for generic $z$ we
could put ${\bf p}_{d-1}=i{\bf p}_{d-1}^E+{\bf p}_{Re}$,
$p_z=ip_z^E+p_{Re}$ and $p_v=i p_v^E-p_{Re}$, what would be a more
general solution. Nevertheless, with this complex momenta, the
imaginary unit $i$ could not be removed from the equations of motion
by Wick rotation of the worldline parameter $\tau=-i\tau_E$, and the
there would be no real-valued Euclidean variables to access the
forbidden region.
As expexcted the square root $R$ in \eqref{emomenta} is real only for spacelike
trajectories. In other words, our Euclidean action \eqref{action4E}
allows us to find trajectories joining spacelike separated boundary
points.

~

The conclusion is that, in the absence of charge, spacelike
separated points in the forbidden region can be joined by real
classical trajectories of the Euclidean worldline time action
\eqref{action4E}, obtained from the original one \eqref{action4} via
the Wick rotation $\tau=-i\tau_E$.  This is a standard prescription and the
one used in \cite{bala}.
%
%
%
%
\subsection{Non-vanishing probe charge}
%
%
%
%
\subsubsection{Wick rotation of the worldline time and the probe charge}
For non-vanishing charge on the other hand, several complications
arise. In this case, the action read
\be
S=\int d\tau \left(-\frac {mL}z\sqrt{-\dot {\bf x}_{d-1}^2+ f
\dot v^2+2\dot z\dot v} +{\sf e}A_v \dot v \right)\,.
\label{action5}
\ee
Writing the momenta \eqref{curvedmomenta} explicitly  we have
\ba p_v&=&- \frac {mL}{zR}(f\dot v+\dot z)+{\sf e}A_v\,,
\nonumber\\
p_z&=& -\frac {mL}{zR}\dot v \,,\qquad\qquad\qquad\qquad {\rm
with}\quad \left\{
\begin{array}{l}
A_v=-{L\gamma\hat Q}z^{d-2}+\mu L\,,
\\ ~\\
R=\sqrt{-\dot {\bf x}_{d-1}^2+f\,\dot v^2+2\dot z\dot v}\,,
\end{array}
\right.
\nonumber\\
{\bf p}_{d-1} &=& \frac {mL}{zR} \dot {\bf x}_{d-1} \,,
\ea
while the relation \eqref{massconstr} becomes
\be fp_z^2-2(p_v-{\sf e}A_v) p_z+{\bf
p}^2_{d-1}=-\frac{m^2L^2}{z^2}\,. \label{rel} \,.
\ee
Again close enough to the boundary the explicit $v$ dependence
disappears, and the momentum $p_v$ becomes a constant. For fixed
$p_v$ the polynomial in $p_z$ in the left hand side of \eqref{rel}
has a minimum at $p_z=p_v-{\sf e}\mu L$ while its right hand side
becomes arbitrarily negative, implying that the relation cannot be
satisfied for real momenta at small $z$. As in the previous case,
this can be solved by pure imaginary momenta ${\bf p}_{d-1}=i{\bf
p}_{d-1}^E$, $p_z=ip_z^E$ and $p_v=i p_v^E$, and we then write
\ba p_v^E&=&i\frac {mL}{zR}(f\dot v+\dot z)-i{\sf e}A_v\,,
\nonumber\\
p_z^E&=&i\frac {mL}{zR}\dot v \,,
\qquad\qquad\qquad\qquad {\rm
with}\quad \left\{
\begin{array}{l}
A_v=-{L\gamma\hat Q}z^{d-2}+\mu L\,,
\\ ~\\
R=\sqrt{-\dot {\bf x}_{d-1}^2+f\,\dot v^2+2\dot z\dot v}\,,
\end{array}
\right.
\nonumber\\
{\bf p}_{d-1}^E &=&- i\frac {mL}{zR} \dot {\bf x}_{d-1} \,,
\ea
and \eqref{massconstr} becomes
\be -f({p^E_z})^2+2(p_v^E+i{\sf e}A_v) p_z^E-({{\bf
p}^E_{d-1}})^2=-\frac{m^2L^2}{z^2}\,. \label{rel4}
\ee
Now, if we want to interpret the momenta as obtained via a Wick
rotation of the wordline time $\tau=-i\tau_E$, we get
\ba p_v^E&=& -\frac {mL}{zR}(f v'+ z')-i{\sf e}A_v\,,
\nonumber\\
p_z^E&=& -\frac {mL}{zR} v' \,,
\qquad\qquad\qquad\qquad {\rm with}\quad
\left\{
\begin{array}{l}
A_v=-{L\gamma\hat Q}z^{d-2}+\mu L\,,
\\ ~\\
R=\sqrt{{{\bf x}'}_{d-1}^2-f\, v'^2-2 z' v'}
\,,
\end{array}
\right.
\nonumber\\
{\bf p}_{d-1}^E &=& \frac {mL}{zR} {\bf x}'_{d-1}\,, \ea
where we see that the imaginary unit multiplying ${\sf e}A_v$ still
avoids a solution of the equations of motion with real coordinates.
In order to allow that, we also perform the analytic continuation
${\sf e}=i{\sf e}_E$, obtaining
\ba p_v^E&=& -\frac {mL}{zR}(f v'+ z')+{\sf e}_EA_v\,,
\nonumber\\
p_z^E&=& -\frac {mL}{zR} v' \,,\qquad\qquad\qquad\qquad {\rm with}\quad
\left\{
\begin{array}{l}
A_v=-{L\gamma\hat Q}z^{d-2}+\mu L\,,
\\ ~\\
R=\sqrt{{{\bf x}'}_{d-1}^2-f\, v'^2-2 z' v'}\,,
\end{array}
\right.
\nonumber\\
{\bf p}_{d-1}^E &=& \frac {mL}{zR} {\bf x}'_{d-1}\,, \ea
These momenta can be obtained from the Wick rotated action
\be
S=\int d\tau_E \left(\frac {mL}z\sqrt{{{\bf x}'}_{d-1}^2- f
v'^2-2 z'v'} +{\sf e}_E A_v v' \right) \,,
\label{action5E}
\ee

In conclusion, by Wick rotating the worldline parameter
$\tau=-i\tau_E$ and the probe charge ${\sf e}=i{\sf e}_E$, we obtain
a mechanical system \eqref{action5E} whose classical solutions with
real valued coordinates are equivalent to the complex-valued
solutions of our original system \eqref{action5} in the forbidden
region.

\subsubsection{Justification by Kaluza-Klein reduction}
A way to understand the previous statement about the Wick rotation
of the probe charge, is by dimensional oxidation of the 
$(d+1)$-dimensional Vaidya metric plus gauge potential to a
purely geometric $(d+2)$-dimensional background. We write
\be ds_{d+2}^2= \frac {L^2}{z^2}\left(-f dv^2 - 2 dv dz + d{\bf
x}_{d-1}^2\right)+ (du+ \kappa A_v dv )^2 \,,
\label{KK}
\ee
where $u$ is the additional spatial direction and $\kappa$ is the
five dimensional gravitational coupling. The Kaluza-Klein reduction
of \eqref{KK} along the direction $u$ corresponds to the
charged Vaidya metric \eqref{RNADSBH}-\eqref{RNADSBHA} we employed
in our calculations.

The action for a massive particle in this $(d+2)$-dimensional 
background is 
\be 
S_{d+2}=-\int d\tau \frac {ML}
 z
\sqrt{ f \dot v^2 + 2 \dot v \dot z - \dot{\bf x}_{d-1}^2 - \frac
{z^2}{L^2}\left(\dot u+
 \kappa A_v \dot v \right)^2
}\,, \label{action5d} 
\ee
whose canonical momenta read
\ba 
p_v &=& -\frac {ML}{zR} \left( f\dot v+\dot z-\frac {z^2 }
{L^2}\left( \dot u+\kappa A_v\dot v\right)\kappa A_v\right)\,,
\nonumber\\
p_z&=& -\frac {ML}{zR}\dot v\,,
\nonumber\\
{\bf p}_{d-1} &=& \frac {ML}{zR}\dot {\bf x}_{d-1}\,,
\quad\qquad\qquad\quad {\rm with}\quad \left\{
\begin{array}{l}
A_v=-{L\gamma\hat Q}z^{d-2}+\mu L\,,
\\ ~\\
R= \sqrt{ f \dot v^2 + 2 \dot v \dot z - \dot{\bf x}_{d-1}^2 -\frac
{z^2}{L^2} \left(\dot u+
 \kappa A_v \dot v \right)^2
}\,,
\end{array}
\right.
\nonumber\\ \nonumber ~\\
p_u &=& \frac {Mz}{LR} \left( \dot u + \kappa A_v\dot v \right)\,,
\nonumber \ea
with the mass shell constraint taking the form 
\be fp_z^2-2(p_v-\kappa p_u A_v) p_z+{\bf p}^2_{d-1}+\frac{p_u^2
L^2}{z^2}=-\frac{M^2L^2}{z^2}\,. \label{rel5d} \ee

In the above equations, the momentum $p_u$ is a conserved quantity.
So, the dynamics of the remaining degrees of freedom can be equivalently described by
making use of the ``Routhian'' ${\cal L}_{4d}$ obtained by Legendre transforming
the original Lagrangian with respect to $u$, namely ${\cal L}_{4d}= {\cal
L}_{5d}-p_u \dot u$. This Routhian plays the role of a Lagrangian
for the remaining coordinates. In other words, the equations of motion for the 
remaining coordinates can be obtained, after fixing the value of $p_u$, 
by varying the action
\be S_{d+1}=\int d\tau \left(-\frac {mL}z \sqrt{f \dot v^2 + 2 \dot
v \dot z - \dot{\bf x}_{d-1}^2}+ {\sf e}A_v\dot v\right)\,,
\label{action6} \ee
where we have defined
\ba m&=& \sqrt{M^2+p_u^2 }\,,
\nonumber\\
{\sf e}&=&\kappa p_u \,.\ea
Action \eqref{action6} is nothing but \eqref{action5} so, as
expected by KK reasoning, the dynamics of the chargeless
$(d+2)$-dimensional particle is equivalent to that of a
$(d+1)$-dimensional charged particle whose mass and charge are
obtained from the $p_u$ momentum.

In \eqref{rel5d} we again notice that for small enough $z$ the
momentum $p_v$ is conserved, and for fixed $p_u$ and $p_v$ the left
hand side of the on-shell relation has a minimum at
$p_z=p_v-kp_uA_v$, while its right hand side goes into arbitrarily
large negative values, implying that it cannot be satisfied with
real momenta. By defining the purely imaginary momenta $p_v=ip_v^E$,
$p_z=ip_z^E$, ${\bf p}_{d-1}=i{\bf p}_{d-1}^E$ and $p_u=ip_u^E$, and
Wick rotating the worldline parameter $\tau = -i \tau_E$, we get
\ba p_v^E   &=& -\frac {ML}{zR} \left( f v'+ z'-\frac {z^2 }
{L^2}\left( u'+\kappa A_v v'\right)\right)\,,
\nonumber\\
p_z^E&=& -\frac {ML}{zR} v'\,,
\nonumber\\
{\bf p}_{d-1}^E &=& \frac {ML}{zR}  {\bf x}'_{d-1}\,,
\quad\qquad\qquad\quad {\rm with}\quad \left\{
\begin{array}{l}
A_v=-{L\gamma\hat Q}z^{d-2}+\mu L\,,
\\ ~\\
R= \sqrt{ -f v'^2 - 2 v'z' + {{\bf x}'}_{d-1}^2 +\frac {z^2}{L^2}
\left(u'+
 \kappa A_v  v' \right)^2
}\,,
\end{array}
\right.
\nonumber\\ \nonumber ~\\
p_u^E &=& \frac {Mz}{LR} \left( u' + \kappa A_v v' \right) \,,\nonumber
\ea
and \eqref{massconstr} now reads
\be
-f{p_z^E}^2+2(p_v^E-\kappa p_u^E A_v) p^E_z-{{\bf
p}^E_{d-1}}^2-\frac{{p_u^E}^2 L^2}{z^2}=-\frac{M^2L^2}{z^2}\,.
\label{rel5dE} \ee
Which can now be fulfilled at small $z$ with real Euclidean momenta.
These equations can be obtained from the $(d+2)$-Euclidean action
\be S_{d+2}^E=\int d\tau_E \,\frac {ML} z \sqrt{ -f v'^2 - 2 v'z' +
 {{\bf x}'}_{d-1}^2
+ \frac{z^2}{L^2}\left(u'+
 \kappa A_v v' \right)^2
} \,.
\label{action5dE} \ee

As above, making use of the conservation of $p_u^E$, the dynamics encoded
in \eqref{action5dE} can be equivalently obtained from the Routhian obtained from
Legendre transform of the Lagrangian \eqref{action5dE}, in other words from
the action
\be S_{d+1}^E=\int d\tau \left(\frac {mL}z \sqrt{-f v'^2 - 2 v' z' +
{{\bf x}'}_{d-1}^2}+ {\sf e}_EA_v\dot v\right)\,, \ee
where we have defined
\ba m&=& \sqrt{M^2-{p^E_u}^2 }\,,
\nonumber\\
{\sf e}_E&=&\kappa p_u^E \,.\ea
So we have re-obtained our action \eqref{action5E} but now by a Wick
rotation of the worldline parameter only in $d+2$ dimensions,
justifying our  $(d+1)$-dimensional procedure of Wick rotating the
worldline parameter and the probe charge.

%
%
%
%
%
\section{WKB approximation}
\label{WKB}
%
%
%
%
We show in this appendix that the geodesic approach \eqref{pin} is equivalent to the WKB approximation 
of the standard Green funcion definition. The standard definition of the holographic Green
function of the boundary operator $\cal O$, dual to a charged scalar field
$\Phi$ in the bulk, follows from the near boundary expansion
\be \Phi= A_+ z^{\Delta_+} + A_- z^{\Delta_-}\,, \label{berp} \ee
where $A_+$ and $A_-$ are functions of $x^\mu$, and then using the
formula
\ba
\langle\mathcal{O}_{\Delta}(x^\alpha_1)\mathcal{O}_{\Delta}(x^\alpha_2)\rangle
&=&\frac{A_+}{A_-}\,, \ea
where $A_-,A_-$ are evaluated in $x^\alpha=x^\alpha_2-x^\alpha_1$.

To obtain the expansion \eqref{berp} we   use the WKB approximation
to solve the Klein-Gordon equation for $\Phi$
\be (g^{\mu\nu}D_\mu D_\nu - m^2)\Phi =0\,. \label{klein} \ee
We start by rewriting
\be 
(\nabla_\mu\nabla^\mu -2i{\sf e}A_\mu \nabla^\mu -i{\sf
e}\nabla_\mu A^\mu -{\sf e}^2A_\mu A^\mu - m^2)\Phi =0\,,
\label{klein2} \ee
and then define $\Phi =\exp(i{\sf S} )$, to get
\be i\nabla_\mu\nabla^\mu  {\sf S} - \nabla_\mu {\sf S}  \nabla^\mu
{\sf S} +(2 q A_\mu\nabla^\mu {\sf S} - iq \nabla_\mu A^\mu)m
-(q^2A_\mu A^\mu +1)m^2=0\,, \label{klein3} \ee
where ${\sf e}=mq$. Next we propose
\be {\sf S} = m S_{1}+ S_0 +{\cal O}\left({\frac 1m }\right)\,, \ee
obtaining to the lowest order
\ba g^{\mu\nu}\left(\partial_\mu S_1  - q  A_\mu\right)
\left(\partial_\nu S_1  - q  A_\nu\right) +1=0\,, \ea
\be i\nabla^\mu(\partial_\nu S_1 - q A_\nu)- 2g^{\mu\nu}\partial_\mu
S_0 (\partial_\nu S_1  - q A_\nu)=0\,. \ee
If we rescale $mS_1=S$ from the first equation we get
\ba g^{\mu\nu}\left(\partial_\mu S  - {\sf e}  A_\mu\right)
\left(\partial_\nu S  - {\sf e}  A_\nu\right) +m^2=0 \label{HJ}\,.
\ea
This is the Hamilton-Jacobi equation for a relativistic spinless
particle. From Hamilton-Jacobi theory, we know that the function $S$
can the identified with the on-shell classical one-particle action, 
its derivatives being the momenta $\partial_\mu S=p_\mu$, which implies that equation \eqref{HJ} is
nothing but the mass shell constraint \eqref{massconstr}
\be g^{\mu\nu}(p_\mu-{\sf e}A_\mu)(p_\nu-{\sf e}A_\nu)+m^2=0\,. \ee
Regarding the second equation above, writing $S_0=-(i/2) \log B^2$,
with $B$ is an arbitrary function, we get
\be \nabla^\mu \left[B^2(p_\mu  - {\sf e}A_\mu) \right] =0
\label{pik}\,.\ee
This equation gives the first quantum correction, and can be
identified with the continuity equation for the probability current
$j_\nu=B^2(p_\nu - {\sf e}A_\nu)$.

 For an asymptotically AdS metric, close to the boundary
$z=0$, there are two approximated solutions of
\eqref{HJ}-\eqref{pik}
\ba S_\pm &\simeq& {\bf p}_{d-1}\cdot {\bf x}_{d-1}-p_v v \pm mL\log
z\,,
\\
B^2&\simeq&\,\frac {c^2 }{z}\,, \ea
depending on integration constants $p_v, {\bf p}_{d-1}$ and
$c$. Extending these solutions $S_\pm$ to the bulk IR region,
we get the general form of the scalar field
\be \Phi=  \frac c{\sqrt{z}} \left(e^{iS_+}+ e^{iS_-}\right)\,, \ee
and we can thus identify
\be A_\pm = \lim_{z\to0}z^{\mp  mL} e^{iS_\pm} B\,. \ee
In other words
\be
\langle\mathcal{O}_{\Delta}(x^\alpha_1)\mathcal{O}_{\Delta}(x^\alpha_2)\rangle=\frac
{A_+}{A_-}= \lim_{z\to0}z^{-2 mL} e^{i(S_+-S_-)}\,. \ee
Now, by noticing that the functions $S_\pm$ correspond to the
classical on shell action integrated from the turning point at the
tip of the trajectory $z=z_*$ into the boundary $z=z_\epsilon$ with
the $\pm$ identifying each of the geodesic branches, we can
establish that $S_+-S_-=S_{\sf on-shell}$. With this, we recover
formula \eqref{correlator}, where we replaced $\Delta = mL$ in the
large $m$ limit (see \cite{hartho} for a related discussion).

%
%
%
%
%
%

\end{document}